\documentclass[bibyear]{aa} 
\usepackage{txfonts}
\usepackage{graphicx} 

\newcommand{\be}{\begin{equation}}
\newcommand{\ee}{\end{equation}}

\begin{document}

\title{Quasi-spherical superclusters}

\author{P. Hein\"{a}m\"{a}ki \inst{1}, P. Teerikorpi \inst{1}, M. Douspis \inst{2}, P.Nurmi \inst{3}, M. Einasto \inst{4}, M. Gramann \inst{4}, J. Nevalainen \inst{4}, Enn Saar \inst{4, 5}}
  \institute{Tuorla Observatory, Department of Physics and Astronomy, Vesilinnantie 5, University of Turku, 20014 Turku, Finland
              \email{pekheina@utu.fi}
         \and
         Institut d’Astrophysique Spatiale, CNRS (UMR8617) Université Paris Sud, Bâtiment 121, Orsay, France
         \and
         Biodiversity unit, University of Turku, 20014 Turku, Finland
         \and
            Tartu Observatory, Tartu University, 61602 T\~oravere, Estonia
            \and 
             Estonian Academy of  Sciences, Kohtu 8, Tallinn, Estonia \\ 
             }

\date{Received / Accepted}

\abstract
{Superclusters are systems with varied properties and varied fractional overdensities.  
Their dynamical state evolves under the influence of two components: dark energy and gravitational force.
The dominant component at any spatial location and cosmic epoch is determined by the total mass and the local overdensity of the system. However, generally the dynamical state of superclusters is poorly known.

}
{We study properties of superclusters and select a sample of quasi-spherical superclusters, the dynamics of which can be studied using the $\Lambda$ significance diagram.   
}
{ 
We extracted our supercluster sample with an adaptive local threshold density method from the Sloan Digital Sky Survey Data Release 7 (SDSS DR7) data and
estimated their masses using the dynamical masses for member galaxies and groups. 
We used topological analysis based on Minkowski functionals and the positions of galaxies and galaxy groups in superclusters. Finally, we highlight the dynamical state of a 
few exceptional types of superclusters found in this study using
the $\Lambda$ significance diagram.  

}
{Our final sample contains 65 superclusters in the distance range of 130 to 450 Mpc. Supercluster masses range between 
 $1.1 \times 10^{15} M_{\sun}$ and $1.4 \times 10^{16} M_{\sun}$ and sizes between 25 Mpc and 87 Mpc.
We find that pancake-type superclusters 
form the low-luminosity, small, poor and low-mass end of superclusters. 
 We find four superclusters of unusual types, exhibiting exceptionally spherical shapes. These so-called quasi-spherical systems contain a high-density core surrounded by a relatively spherical density and galaxy distribution.
The mass-to-light ratio of these quasi-sphericals is higher than those of the other superclusters, suggesting a relatively high dark matter 
content.
Using the $\Lambda$ significance diagram for oblate and prolate spheroids, we find that three quasi-spherical superclusters are gravitationally bound at the present epoch. 
}
{Quasi-spherical superclusters are among the largest gravitationally bound systems found to date, and
form a special class of giant systems that, dynamically, are in between large
gravitationally unbound superclusters and clusters of galaxies in an equilibrium configuration. 
}
 
\keywords{large-scale structure of Universe – galaxies: clusters: general – dark matter – dark energy}

\authorrunning{P. Hein\"{a}m\"{a}ki et al.}
\titlerunning{Quasi-spherical superclusters}
\maketitle

\section{Introduction}

According to the current cosmological paradigm, the great majority of the contents of the Universe 
is governed by the dark sector. In this picture, formation and evolution of the cosmic structure is driven by the competition between the 
gravitational attraction of the dark matter (DM) and the expansion of the space itself. In the 
standard $\Lambda$CDM cosmology, the latter is assumed to be currently accelerated due  
to the constant energy density of the vacuum, which appears as a cosmological constant, $\Lambda,$ in Einstein's field equations.
During the matter-dominated era of cosmic history,
the gravitational attraction of the initial primordial density fluctuations of the DM leads to the formation of increasingly large concentrations of matter. 

While galaxies and clusters are highly concentrated virialized systems with substantial fractional overdensities $\delta_m \sim 10^{6}-10^{2}$, superclusters as a whole are perceived as large and 
relatively loose, unrelaxed agglomerations of galaxies and clusters, with overdensities of just a few 
times the mean density of the Universe. Superclusters are generally surrounded by voids, that is, under-dense regions with $\delta_m  < 1$. 
Due to their small fractional overdensity, superclusters as a whole remained in the linear regime until recent times and finally  
around $z \sim$ 0.5 (Frieman et al. \cite{frieman08}), when the dark energy (DE) density      
started to dominate over the matter density, and the size of the evolving concentrations (the growth of linear perturbations) was halted at supercluster scales. 
On the other hand, because of the non-uniform distribution of gravitating matter, the DE (cosmic `antigravity') can be stronger than gravity
also locally on scales of $\sim 1-10$ Mpc (Chernin \cite{chernin01}; Byrd et al. \cite{byrd12}).
Similarly, we know that superclusters may have collapsing cores and one or several merging clusters in their high-density cores (Einasto et al. \cite{einasto21} and references therein).
Therefore, the presence of DE along with gravitating matter influences the formation of the large-scale structure on all scales from groups of 
galaxies to superclusters (Teerikorpi et al. \cite{teerikorpi15}).

Superclusters form a heterogeneous sample of structures where the observed properties vary considerably in size, richness, luminosity, and 
morphology (Einasto et al. \cite{einasto80}; Davis et al. \cite{davis82}; Einasto et al. \cite{einasto2011}; Liivam\"{a}gi et al. \cite{liivamagi12}).   
Due to the small fractional overdensities of superclusters, their extraction from the observed galaxy surveys requires careful analysis of the galaxy density field.  
Several methods have been introduced for reconstruction of the suitable density field from large galaxy surveys:   
grid-based density with appropriate kernel smoothing (e.g. Basilakos et al. \cite{basilakos01}), luminosity-weighted density field (e.g. Einasto et al. \cite{einasto03}), the Wiener filtering technique (e.g. Erdogdu et al. \cite{erdogdu04}),  and the Voronoi tesselation-based methods (e.g. Nadathur \& Crittenden \cite{nadathur16}). 
A dynamical definition of superclusters was introduced by Tully et al. (\cite{tully14}). 
Using a combination of the density and peculiar velocity fields of galaxies (Cosmicflows-2 catalogue) in our local Universe,  these latter authors defined the Laniakea supercluster  as the volume (with a characteristic diameter of about 160 Mpc) within which the galaxies have an inward peculiar velocity ($v_{pec}$) component.
Outside the outer surface of the Laniakea supercluster, the matter is expanding along with the Hubble flow and velocity ($u=Hr$) or it has 
a peculiar velocity in the direction of some other mass concentrations. Within this concept, the regions of coherent inward 
motions are called `basins of attraction' while the regions of coherent outflow are 
called  `basins of repulsion' (e.g. Dupuy et al. \cite{dupuy}).          
Detailed peculiar velocity information for galaxies is available only for our local Universe. 

To probe the dynamical state of superclusters in much larger cosmological volumes, we must use some other methods. For instance, if one considers only relatively spherical systems, one may use the luminosity density field data together with 
analytical approaches, such as the spherical collapse assumption (e.g. Einasto et al. \cite{einasto20}) and the concept of the $\Lambda$ significance diagram (Teerikorpi et al. \cite{teerikorpi15}) to characterise the dynamical 
state of a supercluster or its parts   
(e.g. Gramann et al. \cite{gramann15}; Einasto et al. \cite{einasto15}; Chon et al. \cite{chon15}, Teerikorpi et al. \cite{teerikorpi15}; Einasto et al. \cite{einasto21}). The $\Lambda$ significance diagram presents a useful way to distinguish dynamically different (expanding or contracting) regions from each other on different spatial and mass scales around superclusters.   
In the dynamical picture of structures, at one end we have large 
gravitationally unbound superclusters defined as systems with inward peculiar velocity components 
relative to the  Hubble flow, and at the other end there are clusters of galaxies in an equilibrium configuration. It is expected 
that between these extremes there is a continuum 
of structures with different dynamical states.

In this paper, our aim is to use topological tools 
and the $\Lambda$ significance diagram in order to study the properties and dynamical state of the superclusters from the Sloan Digital Sky Survey Data Release 7
(SDSS DR7) sample. To achieve the latter objective,
we are especially interested in the most spherical superclusters of the sample. 
We use the Hubble constant $H_0$ = 67 km$^{-1}$ Mpc$^{-1}$, and for the global density of the dark energy, $\rho_{\Lambda} = 6\times 10^{-30}$ g cm$^{-3}$ (Planck Collaboration XVI \cite{planck14}). The dimensionless Hubble parameter h=0.67 is used in all calculations made in this paper.

\section{The sample of SDSS superclusters}
\subsection{Data}

We use the flux-limited supercluster sample drawn from the SDSS DR7 main survey by Liivam\"{a}gi et al. (\cite{liivamagi12}). This was constructed using the 3D luminosity 
density field of galaxies smoothed with a 8 $h^{-1}$ Mpc kernel. 
Before compiling the supercluster catalogue, galaxy data were reduced  
to minimise various selection effects.
First, to calculate the co-moving distances of the galaxies, their redshifts were corrected into the CMB frame. Secondly, the SDSS sample is very narrow at small distances, and becomes very
diluted at the far end of the sample. Therefore, we selected the lower and upper distance limits
of the sample to be z = 0.009 and z = 0.2. Finally, the absolute magnitudes of galaxies were calculated taking into account galactic extinction and k+e-corrections.

As a next step, groups and clusters of galaxies were determined using a modified friends-of-friends (FoF) algorithm described in detail in Tago et al. \cite{tago10}. In the redshift space, both radial and transverse linking lengths were determined 
separately.  Due to the use 
of a flux-limited sample, the linking length was
increased slightly as a function of distance
(for more details, see Tago et al. \cite{tago10}).
Distances based on spectroscopic redshifts are affected by peculiar velocities of galaxies in groups and clusters. 
This effect of the radial redshift distortions (the finger-of-god effect) on the distances was corrected using the standard deviation $\sigma_{r}$ of the projected galaxy distances from the group centre and the standard deviation $\sigma_{v}$ of the radial velocities of galaxies (both in physical coordinates at the group location). The corrected radial distance of each galaxy of the group was calculated using the ratio of the mentioned standard deviations multiplied by the distance between the group centre and the galaxy. This procedure efficiently suppresses the apparent elongation of groups in the redshift space (for more details, see Liivam\"{a}gi et al. \cite{liivamagi12}, Tempel et al. \cite{tempel12}). 
The luminosity density field calculations are carried out in the Cartesian grid. For that purpose, the corrected galaxy distances were also transferred from the SDSS angular coordinates to the Cartesian coordinates.

Luminosities of the flux-limited sample of galaxies were corrected for the distance selection effect using the procedure by Tempel et al. (\cite{tempel11}). Here the luminosity of each galaxy was corrected by multiplying it by a distance-dependent weighting factor. This factor is defined as the ratio of the expected total luminosity to the luminosity within the magnitude limits at the distance of the galaxy. The final luminosity density field was calculated using a $B_3$ spline kernel of the scale of 8 $h^{-1}$ Mpc and the corrected luminosities (see more details in Liivam\"{a}gi et al. \cite{liivamagi12}). 

The final sample of superclusters (a total of 1313) was determined from the luminosity density field using an $\emph{adaptive density threshold method}$ which gives each supercluster 
its own local threshold limit.
It is important to note that unlike in the ordinary method based on fixed density levels, in the adaptive method 
every supercluster has its own limiting density level, similarly to other astronomical objects (see Liivam\"{a}gi et al. \cite{liivamagi12} for details). 
As a result, many superclusters defined using the fixed density level method
are split into several superclusters in the adaptive case. 
Thus, the number of superclusters is 
higher ($\sim 10\%$) in the adaptive method catalogue. 

In addition to total luminosities (the total weighted luminosity of the galaxies and galaxy groups) of the superclusters, this catalogue gives for example the maximum 
sizes and volumes of superclusters. The volume of a supercluster is calculated from the luminosity density field as the summed volumes of the connected 1 ($h^{-1}$ Mpc)$^3$ grid cells 
(the resolution of the density field calculation) within this supercluster. 
This method does not assume any specific shape for the structure, making morphological analysis 
possible.
The catalogue gives morphology information for the superclusters based on Minkowski functionals and shapefinders derived from them.  

\subsection{Masses of superclusters and selection biases}

The original catalogue by Liivam\"{a}gi (\cite{liivamagi12}) did not contain
mass estimates for superclusters.
In this study, we approximate their masses using dynamical masses.
The dynamical mass is obtained by summing together the dynamical masses of all the galaxy groups belonging to a supercluster.
For that purpose, we first identified galaxy groups of Tago et al. (\cite{tago10}) in our superclusters.
We estimate the dynamical masses of the groups within the radius where the average
density is 200 times the critical density($\rho_c$) using Eq. $8$ in Tempel et al. (\cite{tempel14}):
\be
M_{200}^D = 7.0\times 10^{12} \frac{R_{\rm g}}{{\rm Mpc}}(\frac{\sigma_{\rm v}}{100 {\rm km/s}})^2 M_{\sun} ,\centering
\ee
where $\sigma_{\rm v}$ is the 1D velocity dispersion and the gravitational radius $R_{\rm g}=4.582 \sigma{\rm_{sky}}$. According to Tempel et al. (\cite{tempel14}), the velocity dispersion is estimated
using the line-of-sight velocities of all detected galaxies of
a galaxy group (or cluster) following the standard
formula $\sigma_{v}^2=((1+z_m)^2(n-1))^{-1}\sum_{n=1}^{n} (v_i-v_{mean})^2$, 
where $v_{mean}$ and $z_m$ are the mean group velocity and redshift,
respectively, $v_i$ is the velocity of an individual group member,
and n is the number of galaxies with observed velocities within
the group.  Assuming dynamical symmetry,
the real (3D) velocity dispersion in groups would be
$\sigma_{v}=\sqrt{3}\sigma_{v1D}$.
In order to estimate $R_g$  of the galaxy groups, the rms deviation $\sigma_{sky}$ of
the projected distances of group galaxies are calculated using the formula
$\sigma_{sky}^2=(2n(1+z_m)^2)^{-1}\sum_{n=1}^{n}(r_i)^2$, where $r_i$ is
the projected distance in the sky from the group centre (in comoving coordinates).
The equation 
of the gravitational radius $R_{\rm g}$ 
comes from mass estimations based on the Hernquist profile, where 
$\sigma{\rm_{sky}}$ is the rms deviation of the projected distances of group galaxies in the sky from the group centre. Within the mass range 
$10^{9}{h^{-1} \,M_\sun}$ -- $10^{15}{h^{-1} \,M_\sun}$, the Hernquist masses are 
1.55 -- 1.75 times higher than the NFW masses. We use the average value 1.67 to convert the masses to approximate the NFW masses, which are commonly used estimates of
the dark matter halo masses.
For the group dynamical masses $\geq 10^{14} \,M_{\sun}$, the errors are estimated using the results based on the cluster mass reconstruction analysis by 
Old et al. (\cite{Old13}, \cite{Old15}), where over 20 different cluster-mass-estimation techniques were analysed (including the method used here) aiming to study
statistical uncertainties in clusters masses. For the method used here,
they found an intrinsic scatter of 0.3 dex ($\sim$ a factor of 2)  for masses  $\geq 10^{14} \,M_{\sun}$ .
For the less massive $< 10^{14} \,M_{\sun}$ groups, the errors are larger and we used one-sigma confidence level errors for all groups with masses $< 10^{14} \,M_{\sun}$ and richer than five members. 
Old et al. (\cite{Old13}, \cite{Old15}) found in their analysis that, in general, the method used here has a tendency to underestimate dynamical masses.

For small groups ($N < 5$), the observables $\sigma_{\rm v}$ and  $R_{\rm vir}$ are not well defined, which leads to large scatter in the small group masses. 
Following Gramann et al. (\cite{gramann15}) and Einasto et al. (\cite{einasto15}), we assigned to each group with less than five members the median mass of such groups, which in our case 
is $3.3 \times 10^{12} \pm 5.6\times 10^{10} \,{M_\sun}$. 

Each supercluster also contains single galaxies. We assume that, due to the survey magnitude limit, 
every single galaxy is a member of an unresolved small group and therefore
we also assigned to each such galaxy the median mass 
of the groups that have less than five members. 
 In addition, as in
Gramann et al. (\cite{gramann15}), Einasto et al. (\cite{einasto15}) and Einasto et al. (\cite{einasto21}), we assume that the intracluster gas increases the total mass of the supercluster by 10 \%. 
Although the total luminosities of the groups are corrected via a weighting procedure (the ratio of the expected total luminosity
to the expected luminosity in the visibility window) because of the apparent magnitude limit of the sample, the group richness decreases inevitably as a function of distance. 
This means that the virial mass estimation is not equally reliable in the whole range of distances in the SDSS survey. 
Tago et al. (\cite{tago10}) showed that the estimated group richness remains reliable up to a distance
of about 450 Mpc. After applying this limit, the number of superclusters in the original sample was reduced to 163 superclusters.

Einasto et al. (\cite{einasto2011}) estimated that the minimum richness (number of galaxies in the supercluster) to resolve the morphological properties of superclusters is near the value 300. We also used Monte-Carlo simulations to test the reliability of the correlation of the Minkowski shape parameters and 
the galaxy number density in superclusters and found an agreement with  
the requirement of the minimum richness ($\sim$ 300).
Based on this, and in order to maximise the number of the superclusters in our analysis, we required that a supercluster richness be more than 280 galaxies.

\begin{figure}[htbp] 
\centering
\includegraphics[height=7cm,angle=0]{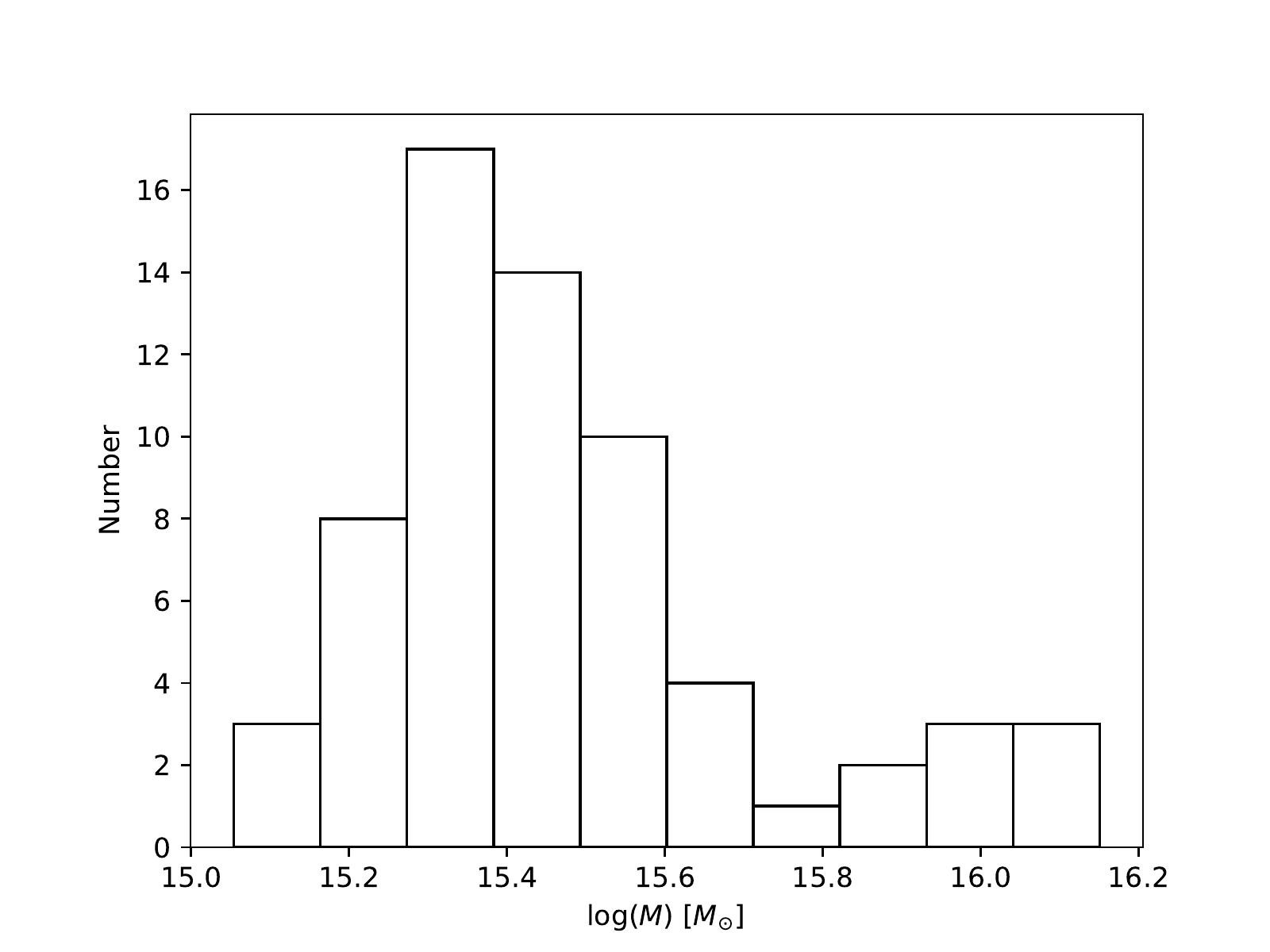}
\resizebox{9cm}{!}{\includegraphics[width = 3.0in,angle=0]{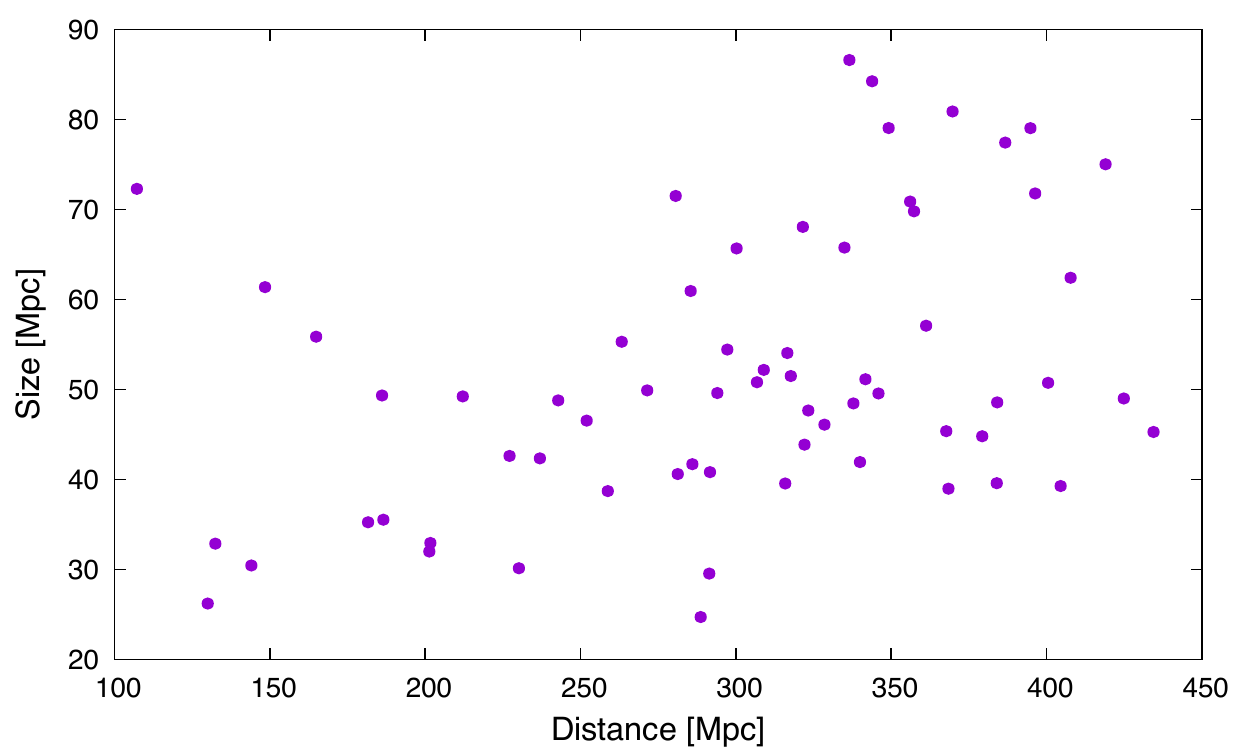}} 
\caption{Mass distribution and size--distance scatter plot of the superclusters. The upper figure shows the mass histogram of the final sample of the 65 superclusters and the lower figure shows supercluster diameter as a function of distance.}
\label{FigMassSC}
\end{figure}

Finally, we removed all the superclusters that have a contact with the edge of the survey mask.  
The original sample contains 1206 superclusters, and after 
applying the selection criteria the final sample for our analysis contains 65 superclusters. 
Figure \ref{FigMassSC} shows the mass distribution of the final supercluster sample. 
The supercluster masses range between $1.1 \times 10^{15} M_{\sun}$ and $1.4 \times 10^{16} M_{\sun}$. The median mass is $2.6 \times 10^{15}$. The lower panel in Fig. \ref{FigMassSC} shows the supercluster size (diameter) distribution as a function of the co-moving distance. Size is defined as a maximum distance between the galaxies inside a supercluster. Sizes range between 25 Mpc and 87 Mpc, while the median size is about 49 Mpc. The figure shows 
a large amount of scatter for supercluster size as a function of distance. Due to the survey geometry, the number of superclusters increases with distance as the survey volume increases. Our result agrees with those of Liivamägi et al. (\cite{liivamagi12}), who showed that the average supercluster size 
remains approximately the same across the survey volume.
The gap in the distribution around 150 -- 250 Mpc is the known 
void region between the nearby superclusters (Einasto et al. \cite{einasto2011}).

The masses of the large well-known superclusters have been studied by several authors. 
The mass of the Shapley supercluster was estimated by Sheth \& Diaferio (\cite{Sheth}) and Ragone et al. (\cite{Ragone}) for example, who found values of $1-2 \times 10^{16} M_{\sun}$. 
Einasto et al. (\cite{einasto21}, \cite{einasto22}) found that the total mass of the 
Corona Borealis supercluster and the superclusters in the 
Sloan Great Wall and in the BOSS Great Wall lie in the range of 
$0.47-1 \times 10^{16} M_{\sun}$.
Recently, Böhringer \& Chon (\cite{Boh21}) used X-ray-luminous galaxy groups and clusters to estimate the masses of eight superclusters. The four most prominent of them (the Perseus-Pisces, the Centaurus, the Coma, and the Hercules supercluster) have masses of between $5 \times 10^{15} M_{\sun}$ and $2.2 \times 10^{16} M_{\sun}$ with sizes from about 40 to more than 100 Mpc. For the four smaller ones, the estimated masses were in the range of $2 \times 10^{15} M_{\sun}$ -- $7 \times 10^{15} M_{\sun}$ with sizes of 20 to 46 Mpc. 
The Laniakea supercluster (Tully et al. \cite{tully14}) has an estimated mass of $\sim 10^{17} M_{\sun}$ and a diameter of about 160 Mpc. We note that in that study, mass and size estimates cover the region of the `basins of attraction', while superclusters in general are extracted using the criterion of fractional overdensity.

 \section{Morphology}

While clusters and groups  are, in general, relatively spherical systems,
the spherical shape obviously is not the usual case for superclusters. Instead,  superclusters most commonly show a filamentary (or elongated prolate) morphology (Jaaniste et al. \cite{jaaniste}; Basilakos \cite{basilakos}; 
Einasto et al. \cite{einasto2011}). 
All adaptively selected superclusters used in our analysis are also elongated at some level.

A useful tool to study supercluster morphology is the shapefinder based on a set of Minkowski functionals
(more details are provided in Sahni et al. \cite{sahni}; Saar et al. \cite{saar2007}; Einasto et al. \cite{einasto07}; Costa-Duarte et al. \cite{costa}). 
Briefly, the final shapefinder is a vector $K=(K_1,K_2)$ whose components K1 and K2 are based on the shapefinders 
$H_i$, (i=1,2,3) as $K_1=\frac{H_1-H_2}{H_2+H_1}$ and $K_2=\frac{H_3-H_2}{H_3+H_2}$.
The shapefinders $H_i$ are determined by 
geometrical 
quantities (V=volume, S=surface area, C=integrated mean curvature) as $H_1=V/S$, $H_2=S/C,$ and $H_3=C$.  
One can use combinations of the shapefinder parameters $K_1$ and $K_2$  to 
measure the shape (planarity, filamentary, sphericity)
of the 3D surface of the structure.

\begin{figure}[htbp] 
\centering
\includegraphics[height=7cm,angle=0]{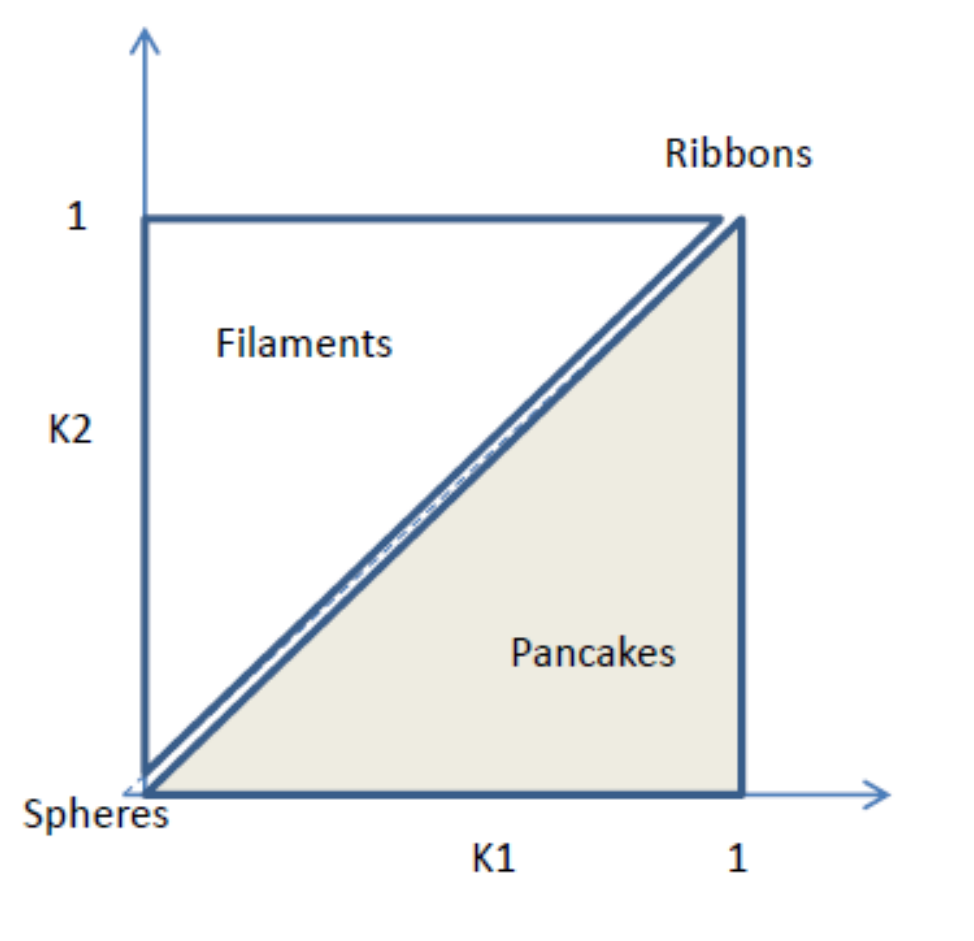}
\caption{Correspondence between the vectors $K_1$ and $K_2$ and different shapes}
\label{FigK1K2}
\end{figure}

 The ratio of the vector components $K_1/K_2 > 1$ corresponds to a `pancake'-type structure (for an ideal pancake $K\approx(1,0)$) while   
$0 < K_1/K_2 < 1$ corresponds to a filament-type structure (for 
an ideal filament $K\approx(0,1)$). When both $K_1$ and $K_2$ are close to zero, the shape is close to a sphere. 
A perfect sphere has the parameters $K_1=0$ and  $K_2=0$.    
If we consider a  triaxial ellipsoid with 
axes $a,b$,and $c$, a sphere $K\approx(0,0)$ is described as $(a,b=a, c=a)$ (Sahni et al. \cite{sahni}; Shandarin et al. \cite{shandarin}). At the other end of the diagonal $K1=K2,$ the combination $K_1=1$ and  $K_2=1$
corresponds to the ribbon-type morphology. Moving along the diagonal towards $K_1=0$ and  $K_2=0,$ the ribbons are filled and grow to spheres.
Figure~\ref{FigK1K2} shows schematically the correspondence between the vector components $K_1$ and $K_2$ and different general shapes.

Though there is no unambiguous correspondence between the values of the shapefinder vectors and certain shapes, the vectors constitute a continuum 
of values which indicate different shapes.
The fourth Minkowski functional $V_3$ describes the sum of the clumps, voids, and tunnels
in the system (e.g. Saar et al. \cite{saar2007}). 
Thus a high value of $V_3$ indicates a complicated clumpy morphology, while $V_3=1$ means a smooth structure with a single centre.

\begin{figure*}[htbp]
 \begin{minipage}{0.24\textwidth}
\resizebox{6cm}{!}{\includegraphics[width = 1.5in,angle=0]{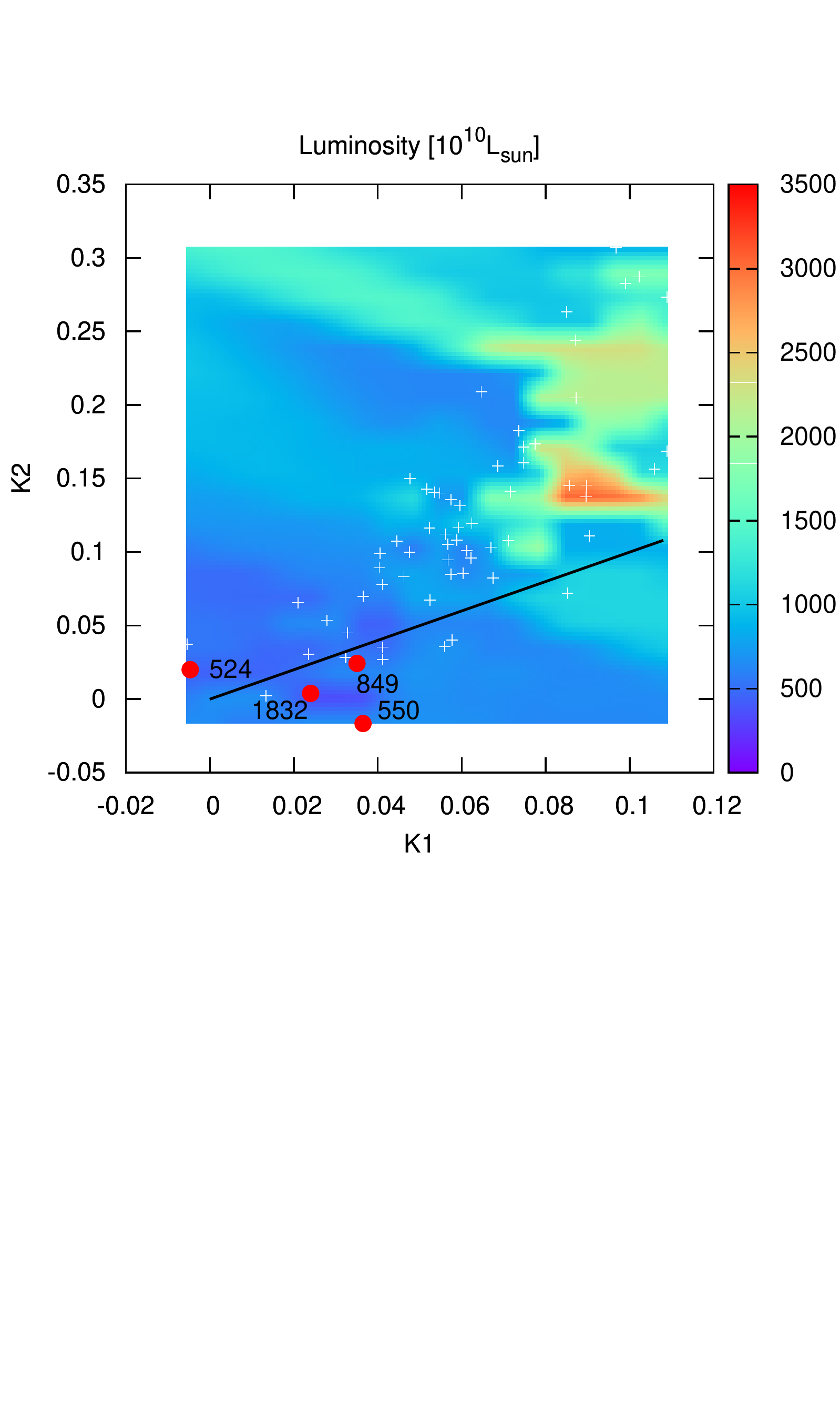}} 
\end{minipage}
\begin{minipage}{0.24\textwidth}
 \hspace*{1.5 cm}
\resizebox{6cm}{!}{\includegraphics[width = 1.5in,angle=0]{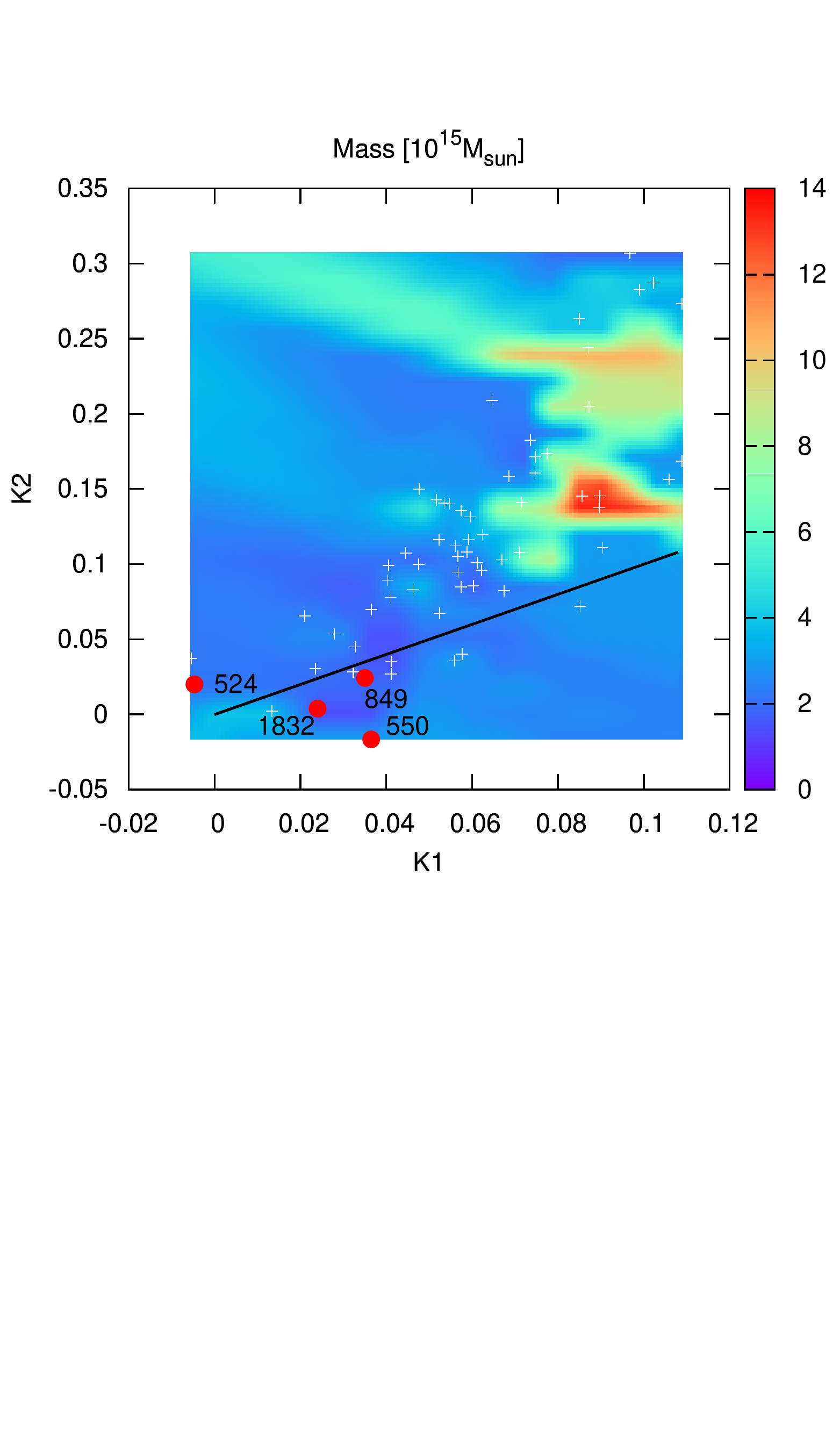}} 
\end{minipage}
\begin{minipage}{0.24\textwidth}
\hspace*{3cm}
\resizebox{6cm}{!}{\includegraphics[width = 1.5in,angle=0,]{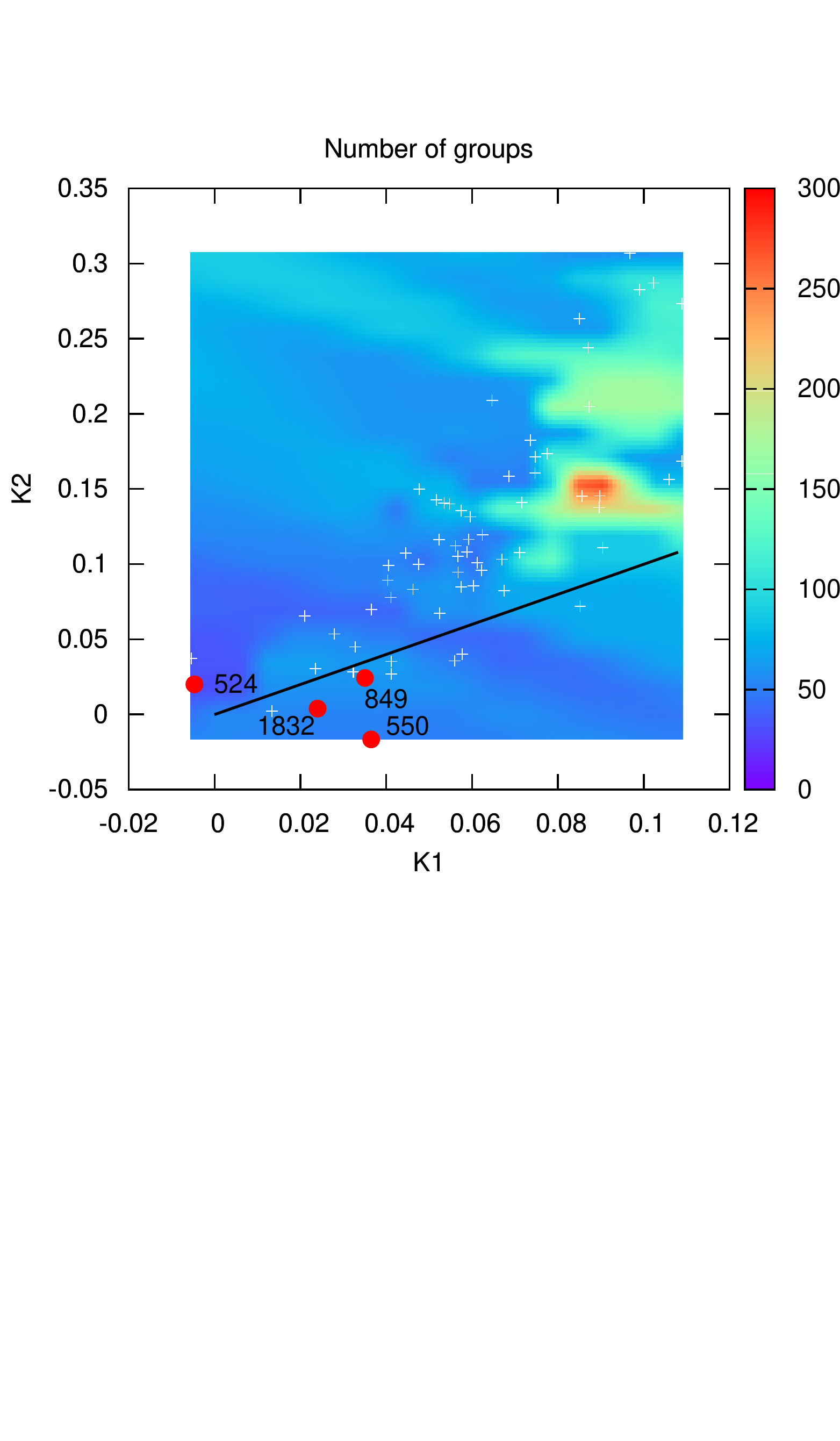}} 
\end{minipage}\\

\vspace*{-4.5cm}

\begin{minipage}{0.24\textwidth}
\resizebox{6cm}{!}{\includegraphics[width = 1.5in,angle=0,]{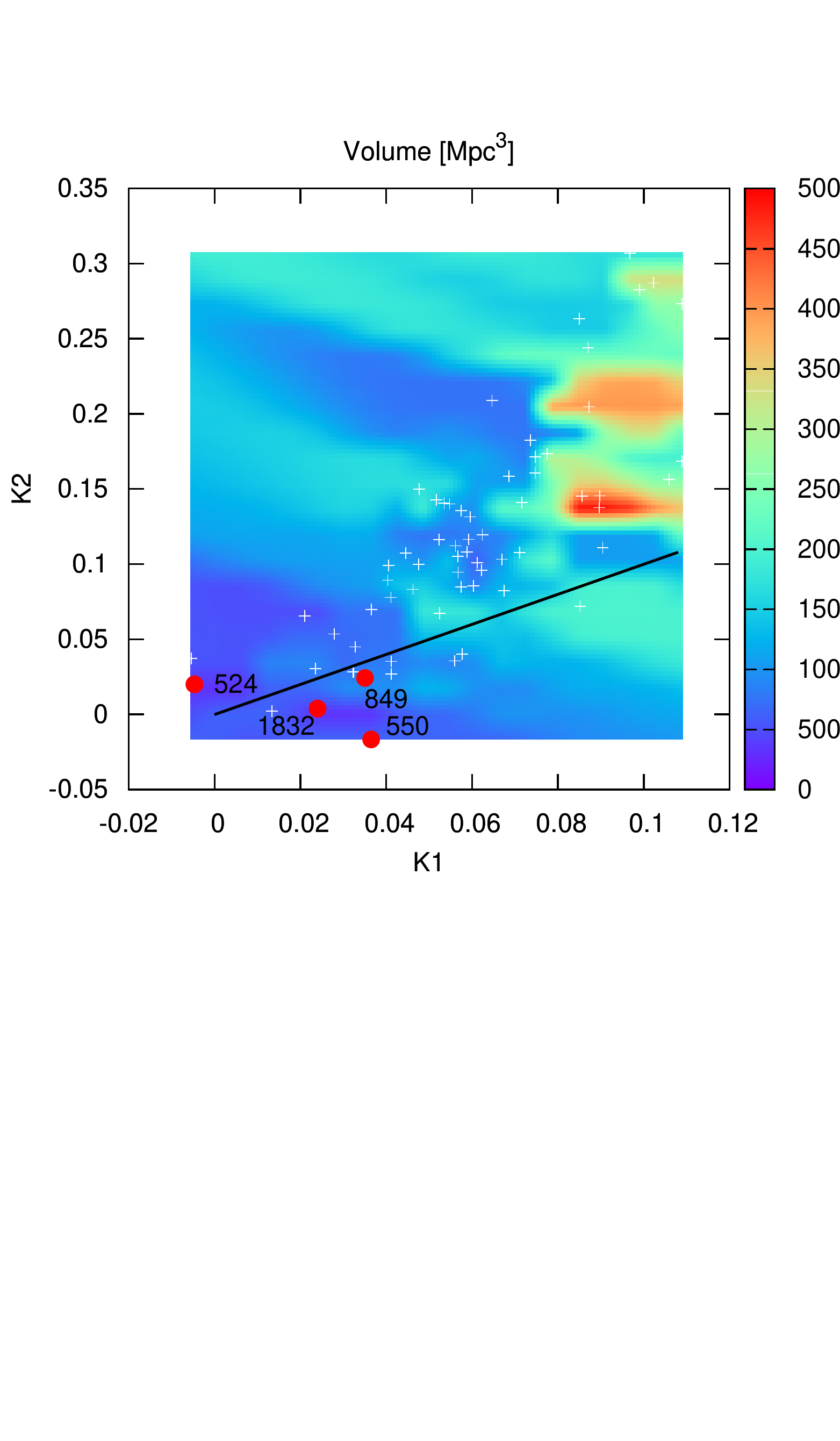}} 
\end{minipage}
\begin{minipage}{0.24\textwidth}
\hspace*{1.5cm}
\resizebox{6cm}{!}{\includegraphics[width = 1.5in,angle=0,]{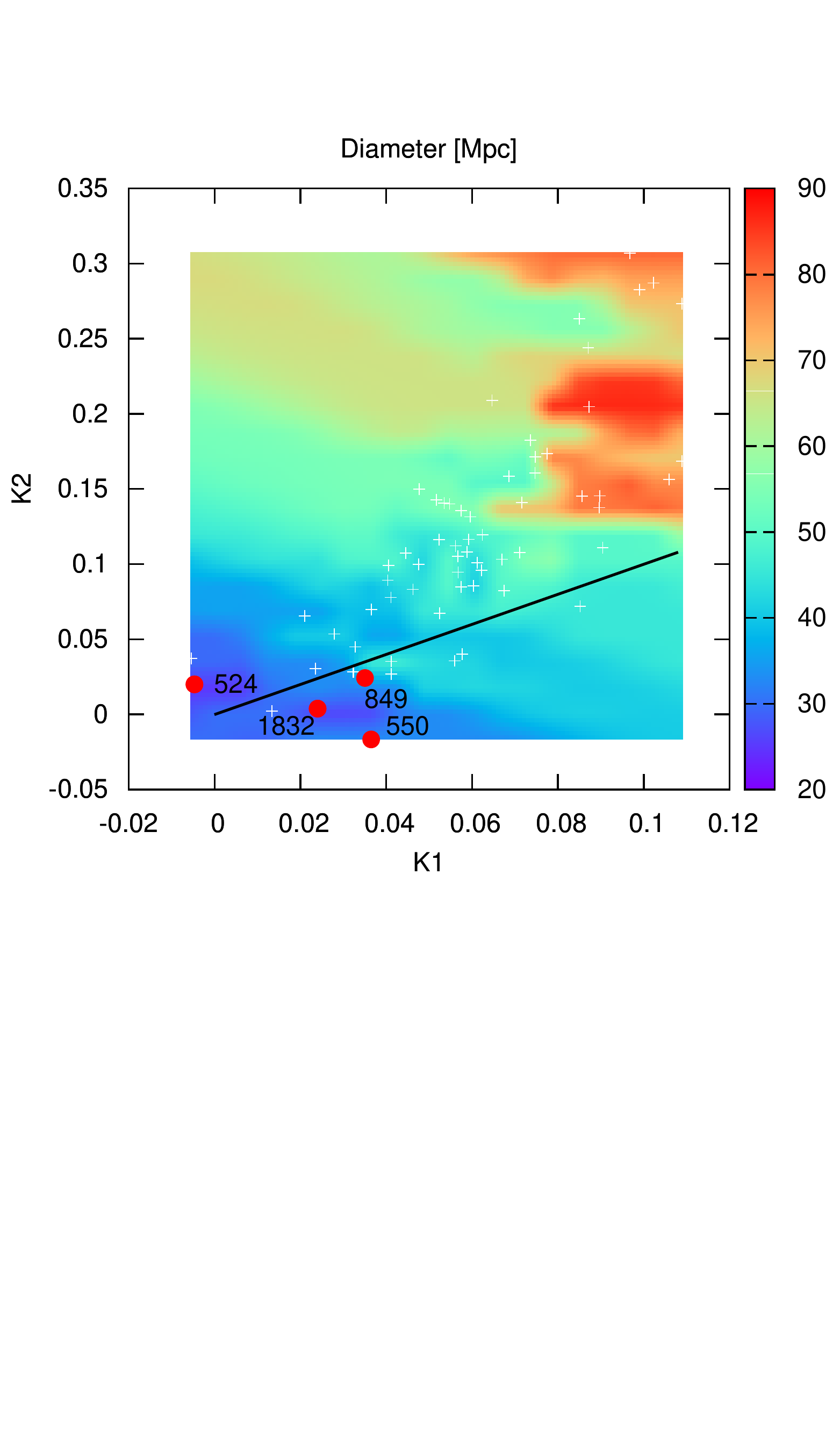}} 
\end{minipage}
\begin{minipage}{0.24\textwidth}
\hspace*{3cm}
\resizebox{6cm}{!}{\includegraphics[width = 1.5in,angle=0,]{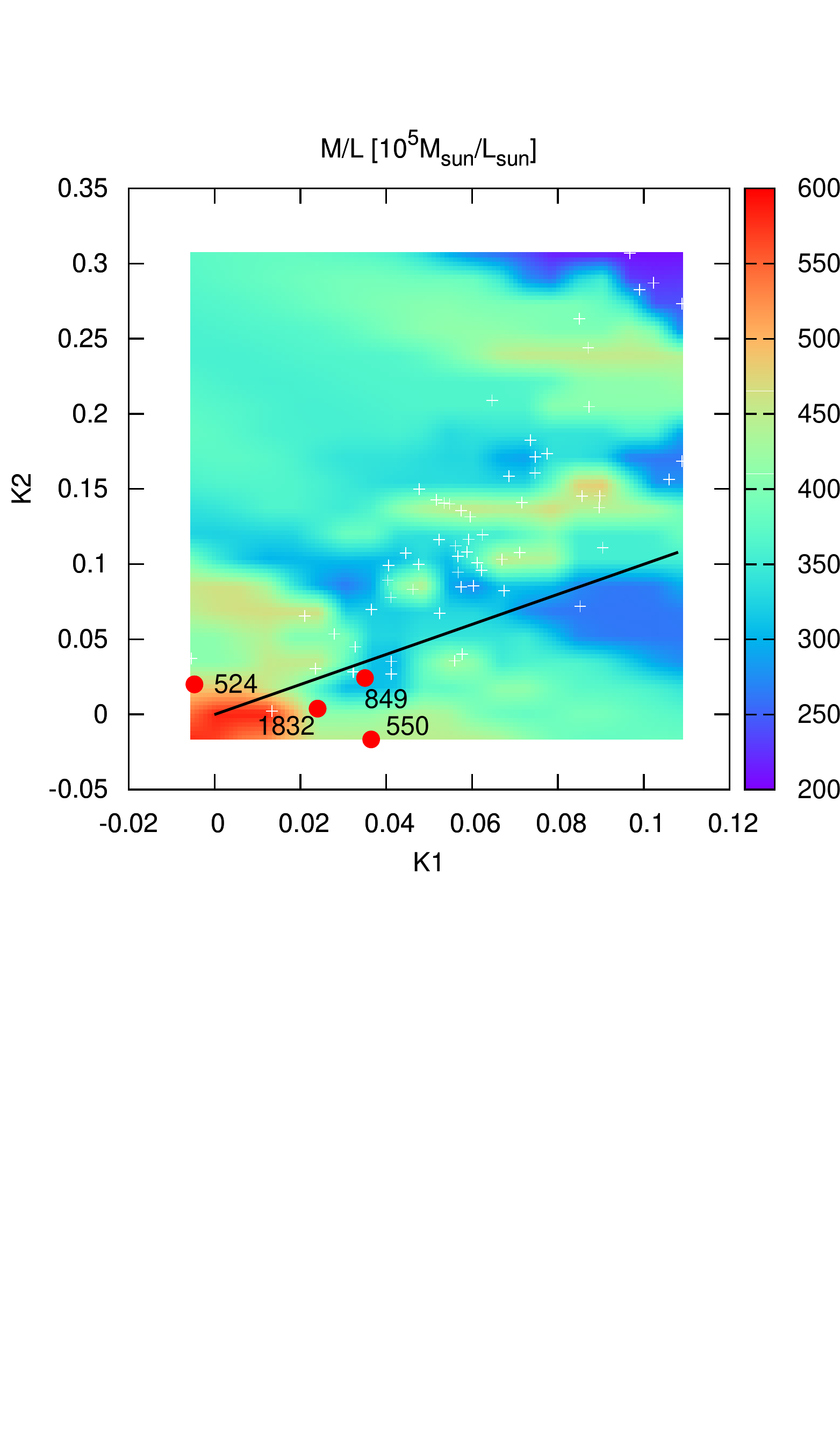}} 
\end{minipage}\\
\vspace*{-3.7cm}
\caption {Coloured contours show the distribution of the values of supercluster properties in the $K1-K2$ plane.    
Quasi-spherical superclusters $K_1$,$K_2$ < (0.05, 0.05) are shown with red points, white points are all the superclusters. Luminosities are in units of $10^{10} L_{\sun}$, masses
in $10^{15}M_\sun$, volumes in Mpc$^{3}$, and diameter in megaparsecs  (Mpc). The dark line indicates $K_1$ = $K_2$  and divides the morphologies between pancakes and filaments.}
\label{FigSCC}
\end{figure*} 

\section {Properties of the superclusters}
 
Figure~\ref{FigSCC} shows the luminosity, mass, number of groups, volume, diameter, and mass-to-light values of the final sample of 65 superclusters smoothed onto the grid on the $K1$ -- $K2$ plane. The distribution of the supercluster properties is illustrated  with coloured contours.
White crosses show the $K1$ -- $K2$ distribution of the final sample of our superclusters.   
Above the black inclined line, the superclusters are of filamentary type and below the line they are of pancake 
type (Fig.~\ref{FigK1K2}). 
  
A majority of the superclusters are of filamentary type (84 \%). 
Among the superclusters, those of filamentary type are more luminous, richer, and larger, in agreement with the  results of Costa-Duarte et al. (\cite{costa}) and Einasto et al. (\cite{E11})  who used 
fixed supercluster threshold limits. Our analysis shows that the dynamical mass follows the same trend. However, 
for the mass-to-light ratio ($M/L$), the division between the filamentary and pancake structures is not as clear. 
   
The red points represent superclusters which have both $K1$ and $K2$ values of less than 0.05. The choice of the limit 0.05 is based on 
our finding that only among superclusters with K1 and K2 
of less than this value can one find systems with a single dominant centre, which is natural for spherical or pancake-type systems. For such systems, the fourth Minkowski functional $V_3=1$. We also required that these systems do not split further at higher density thresholds. To distinguish these superclusters from the other ones, we term them quasi-spherical (QS) superclusters (also referred to hereafter as QSs). 
Figure ~\ref{FigSCC} shows that these QS (a total of four large red dots)
superclusters  are smaller, poorer, less luminous, and less massive than the filamentary-type superclusters, on average.   

In Fig.~\ref{FigSCC} from left to right the red dots represent: QS 524, QS 1832, QS 849, and QS 550. 
For QS 524 $K_1/K_2$< 1, which indicates mild filamentarity.  
In Table~\ref{table1} we present data for these systems.  
Their diameters are between 24 and 33 Mpc and the masses range from $1.2\times 10^{15}$ to $2.7\times 10^{15} \,M_{\odot}$. 

\begin{figure*}[htbp]
\begin{minipage}{0.3\textwidth}
 \resizebox{7.0cm}{!}{\includegraphics[width = 1.0in,angle=0]{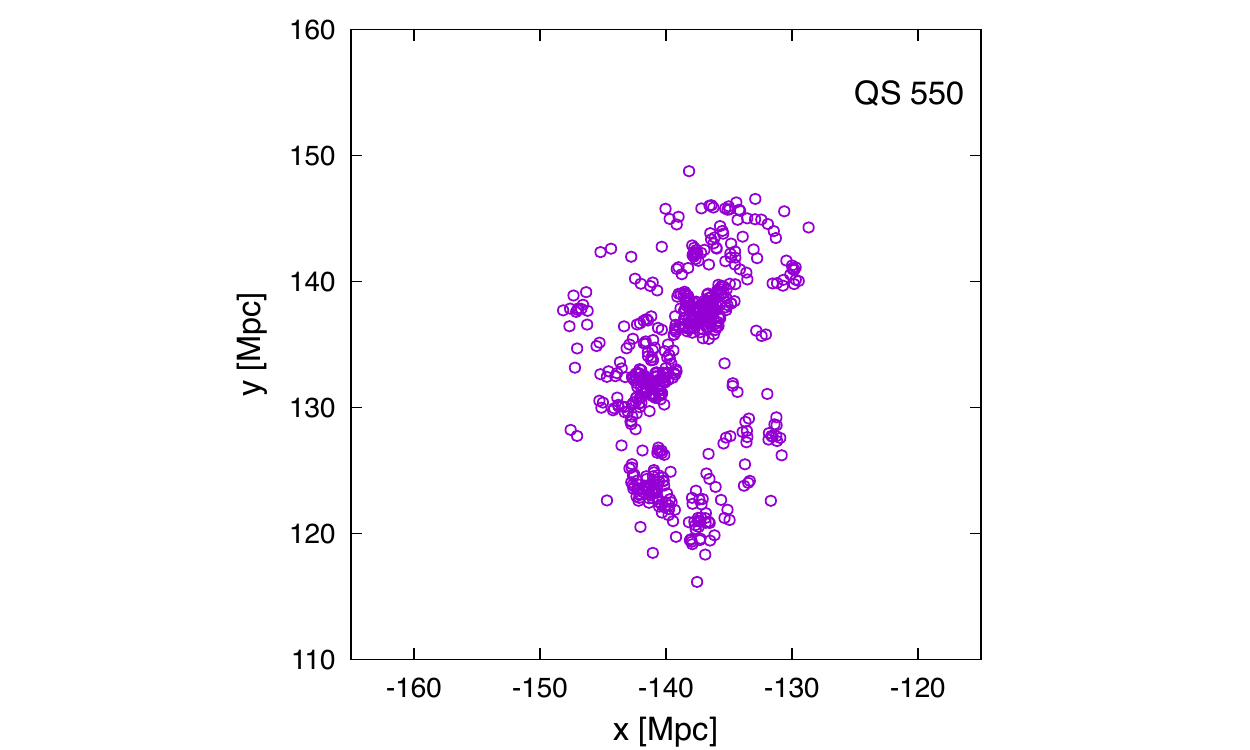}}
\end{minipage}
\begin{minipage}{0.3\textwidth}
\resizebox{7.0cm}{!}{\includegraphics[width = 1.0in,angle=0]{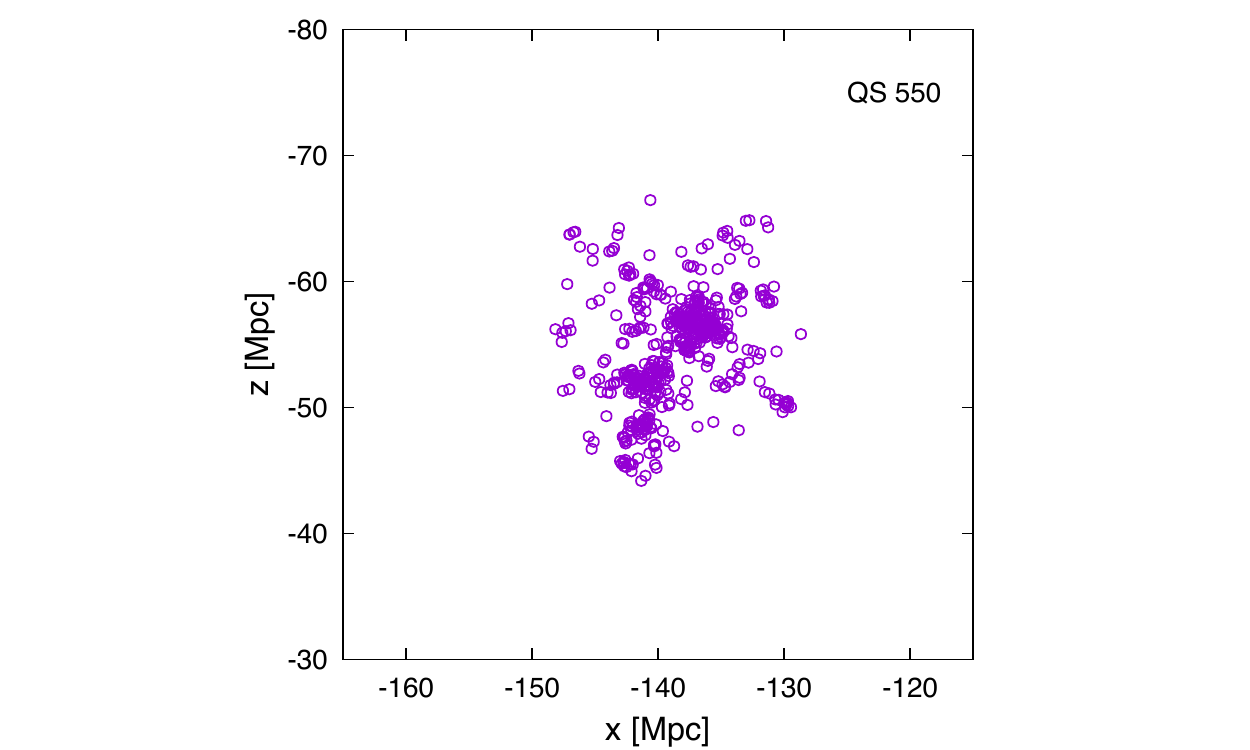}}
 \end{minipage}
\begin{minipage}{0.3\textwidth}
\resizebox{7.0cm}{!}{\includegraphics[width = 1.0in,angle=0]{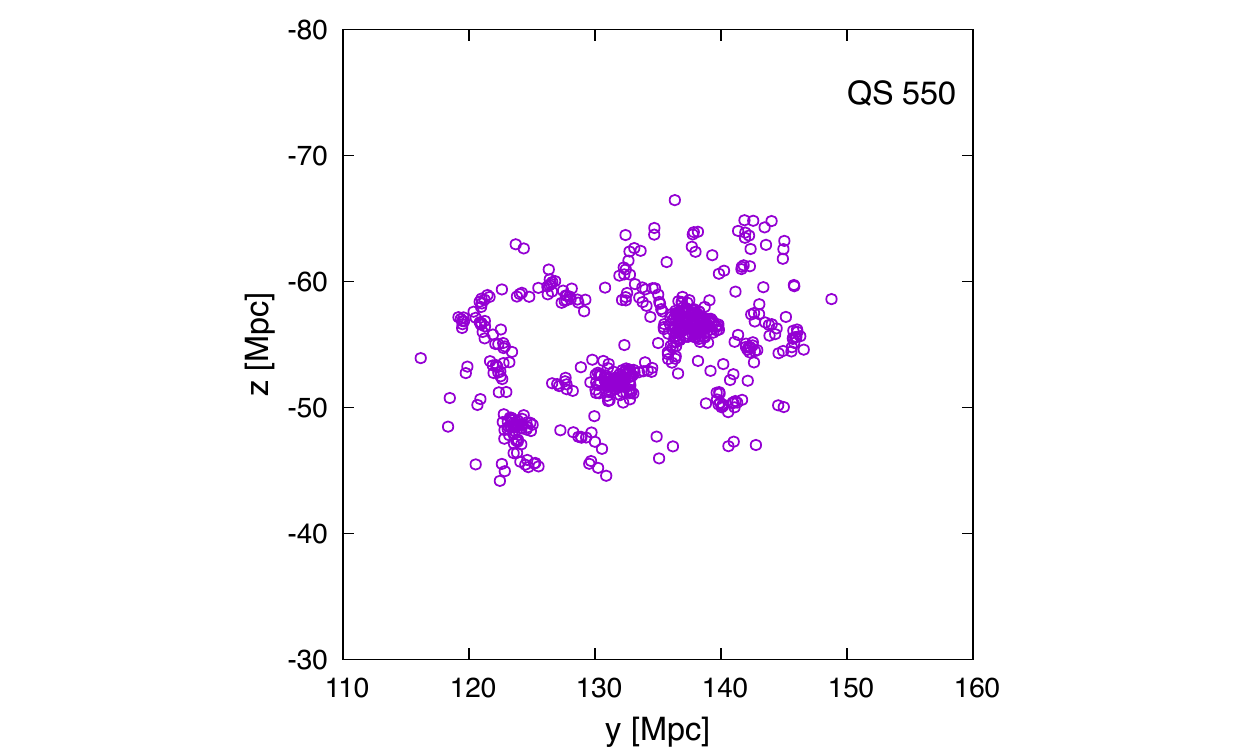}} 
\end{minipage}\\


\begin{minipage}{0.3\textwidth}
\resizebox{7.0cm}{!}{\includegraphics[width = 1.0in,angle=0]{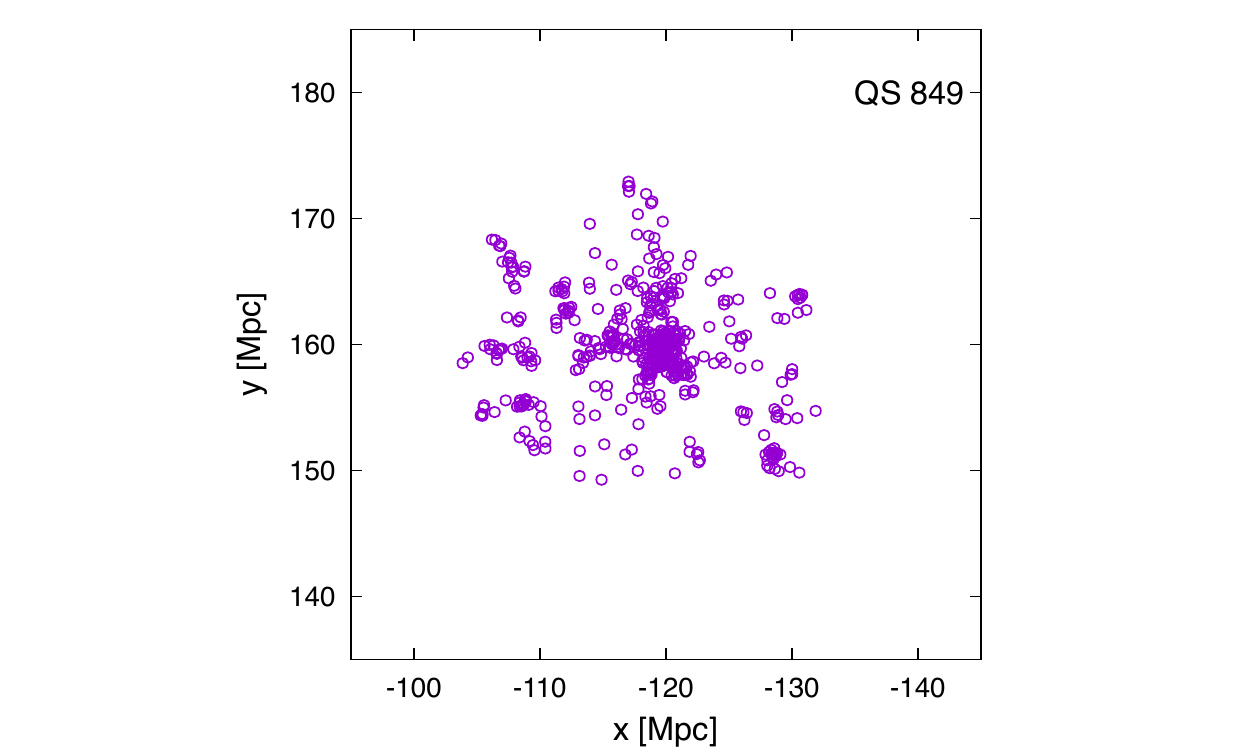}}
\end{minipage}
\begin{minipage}{0.3\textwidth}
\resizebox{7.0cm}{!}{\includegraphics[width = 1.0in,angle=0]{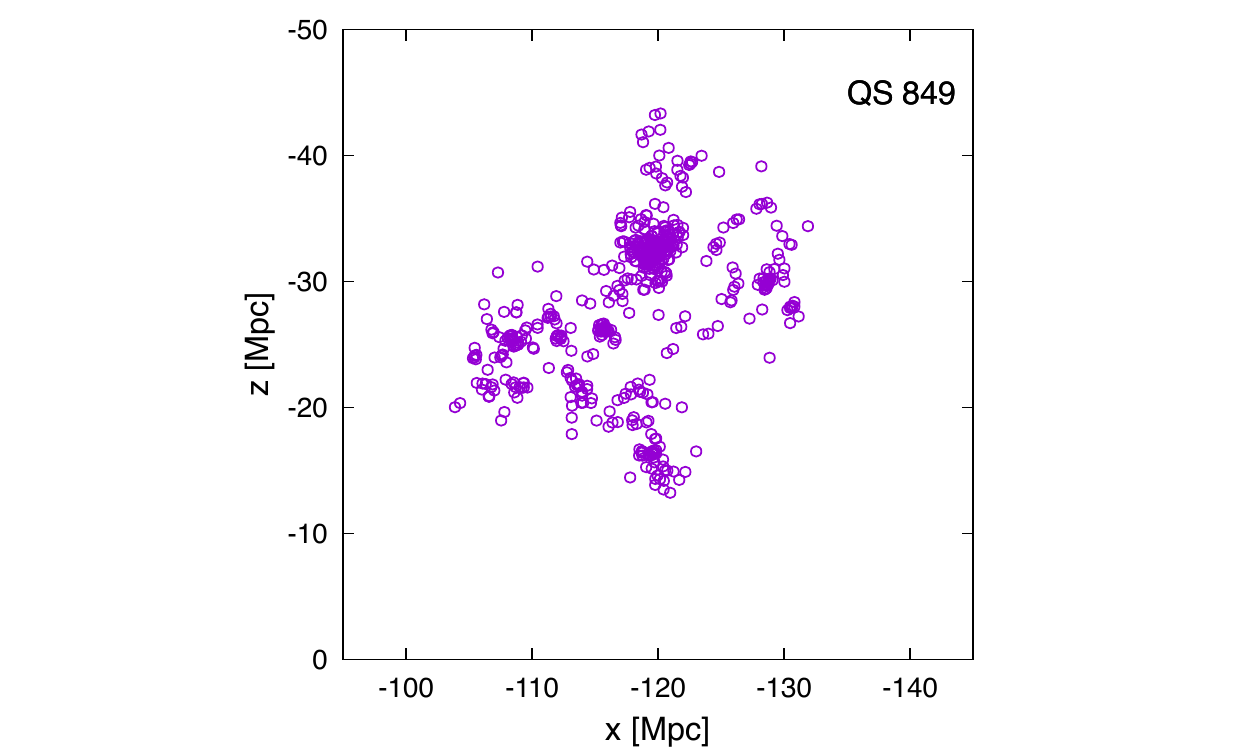}} 
\end{minipage}
\begin{minipage}{0.3\textwidth}
\resizebox{7.0cm}{!}{\includegraphics[width = 1.0in,angle=0]{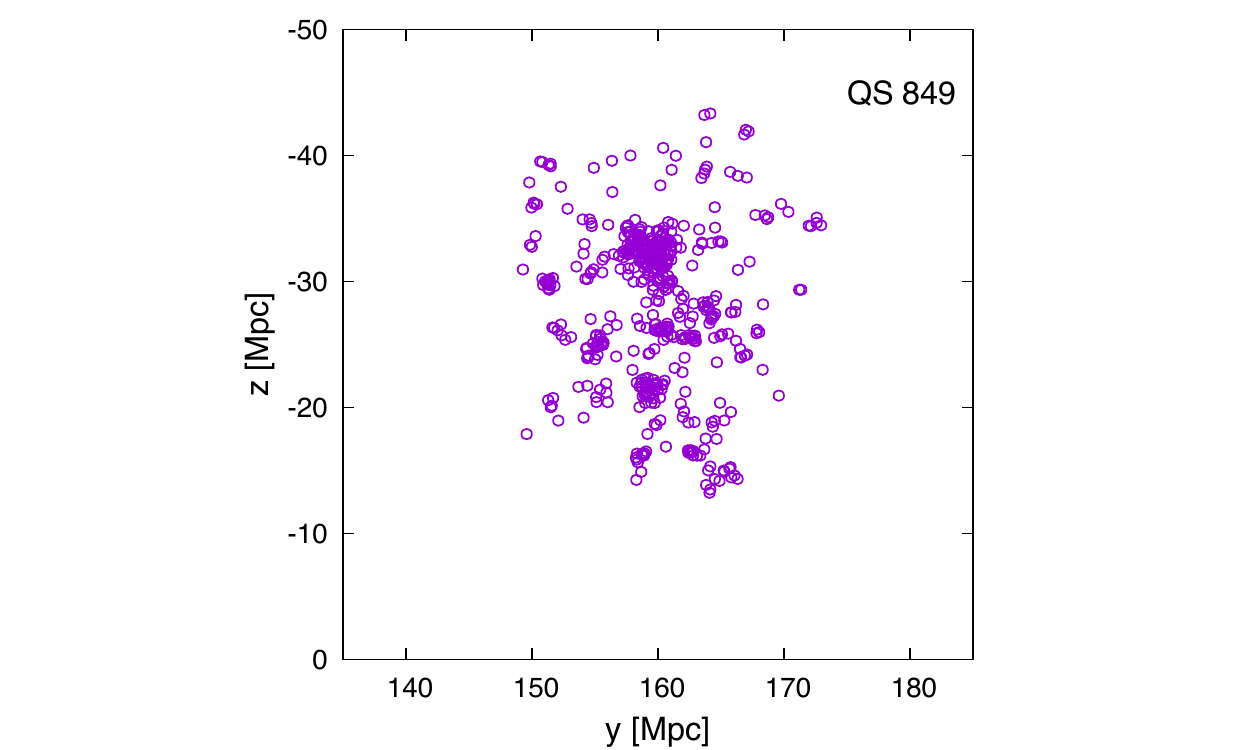}}
\end{minipage}\\


\begin{minipage}{0.3\textwidth}
\resizebox{7.0cm}{!}{\includegraphics[width = 1.0in,angle=0]{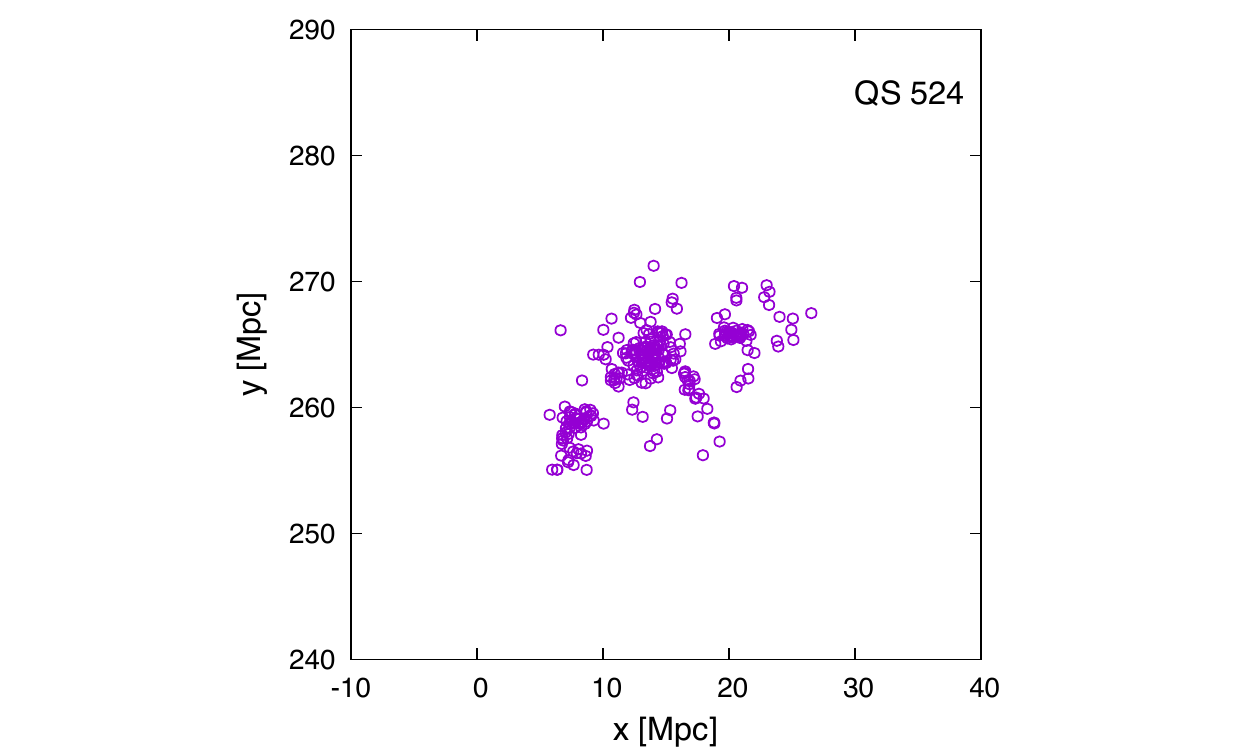}}
\end{minipage}
\begin{minipage}{0.3\textwidth}
\resizebox{7.0cm}{!}{\includegraphics[width = 1.0in,angle=0]{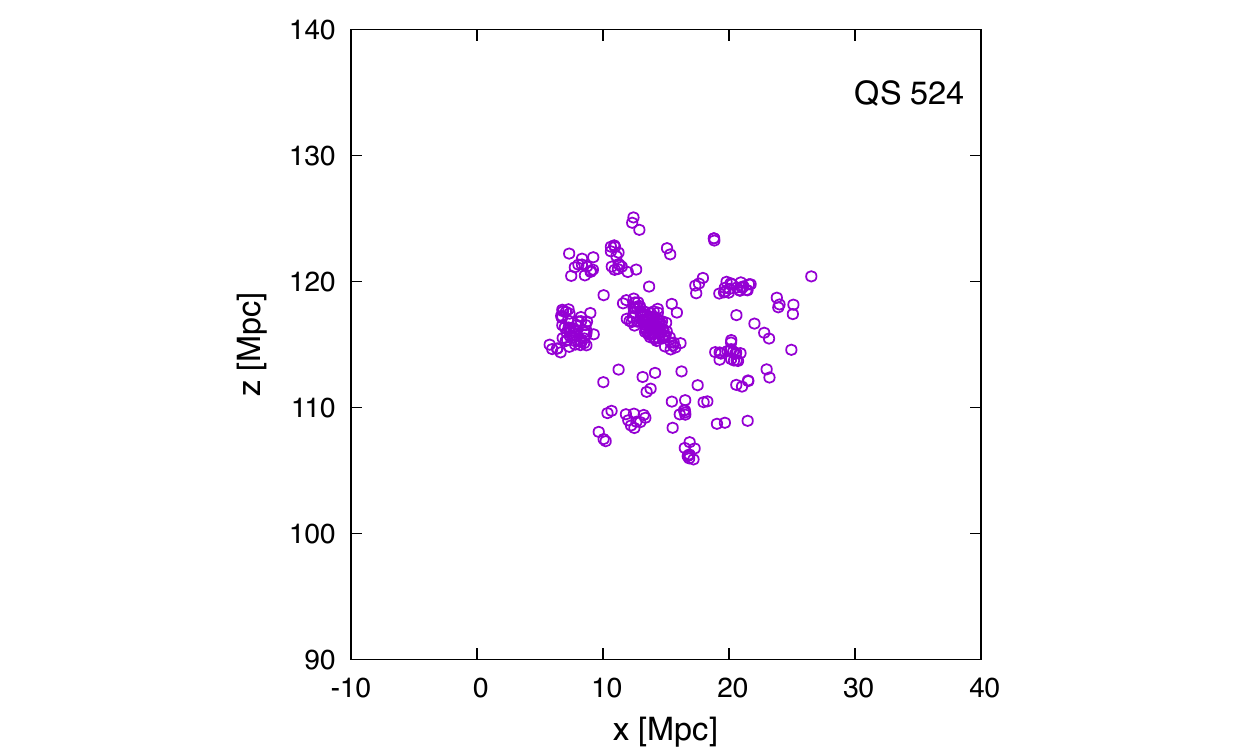}} 
\end{minipage}
\begin{minipage}{0.3\textwidth}
\resizebox{7.0cm}{!}{\includegraphics[width = 1.0in,angle=0]{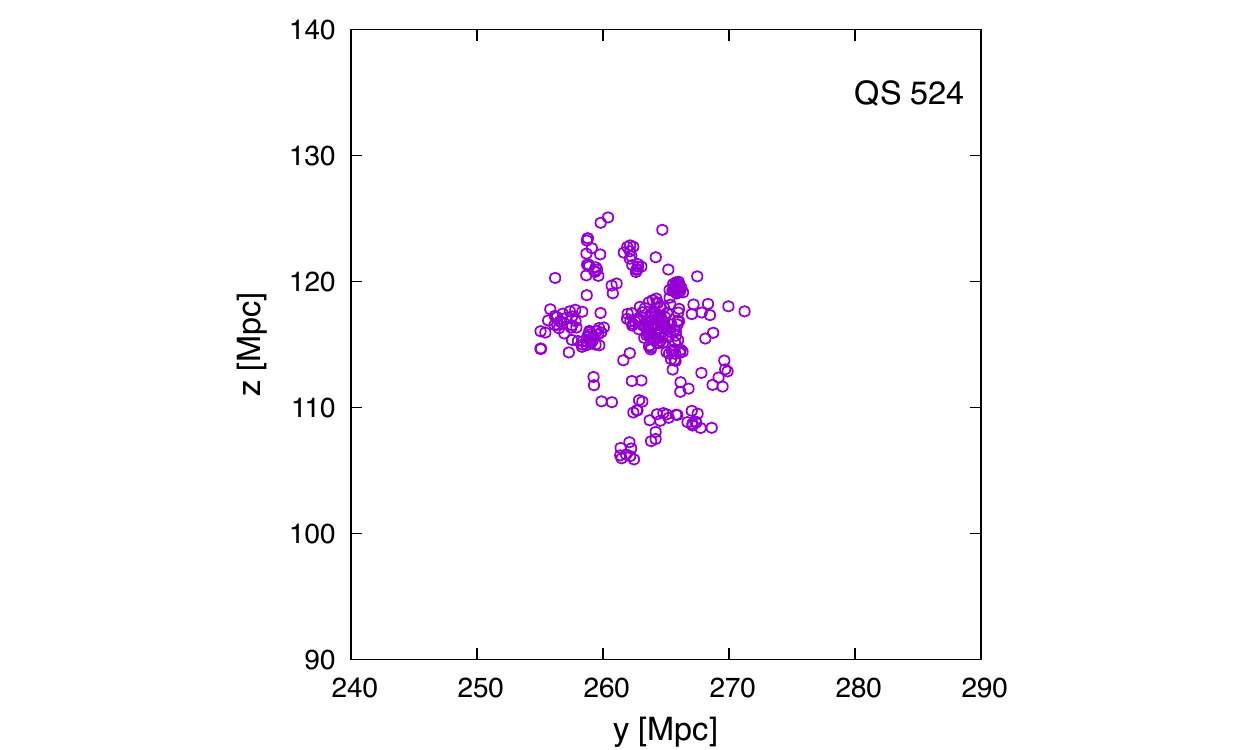}}
\end{minipage}\\


\begin{minipage}{0.3\textwidth}
\resizebox{7.0cm}{!}{\includegraphics[width = 1.0in,angle=0]{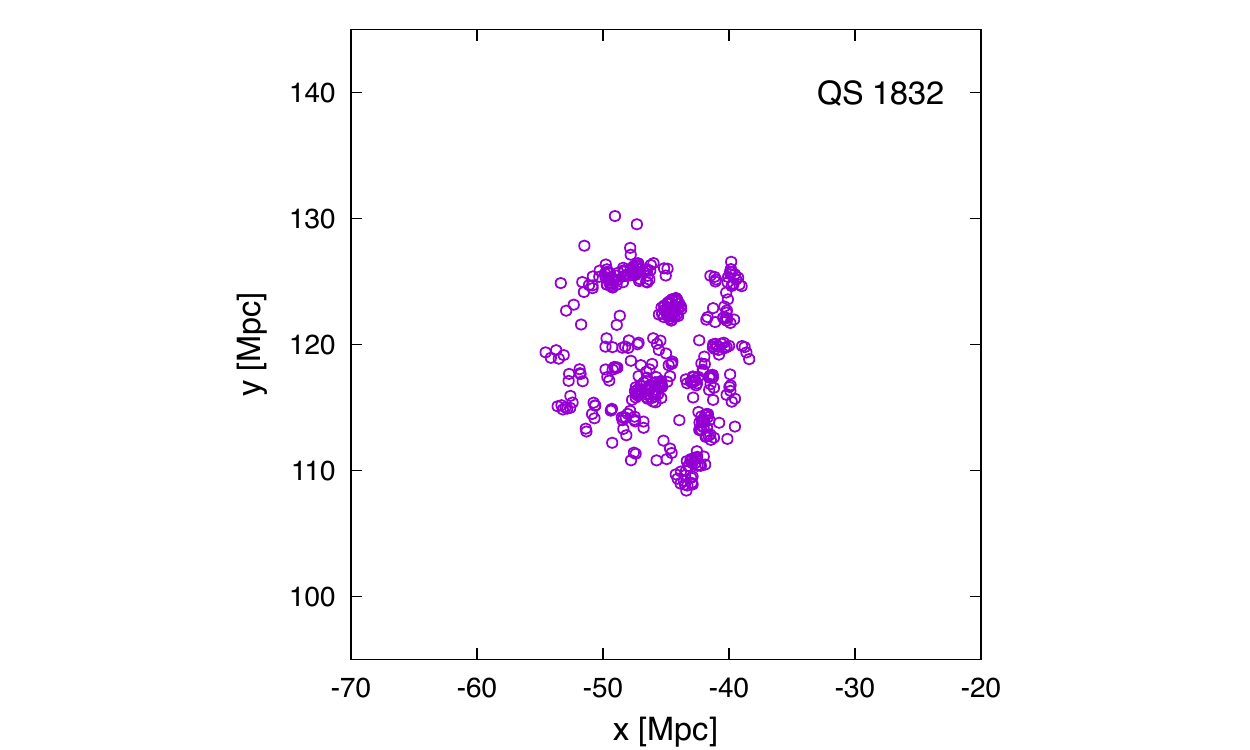}}
\end{minipage}
\begin{minipage}{0.3\textwidth}
\resizebox{7.0cm}{!}{\includegraphics[width = 1.0in,angle=0]{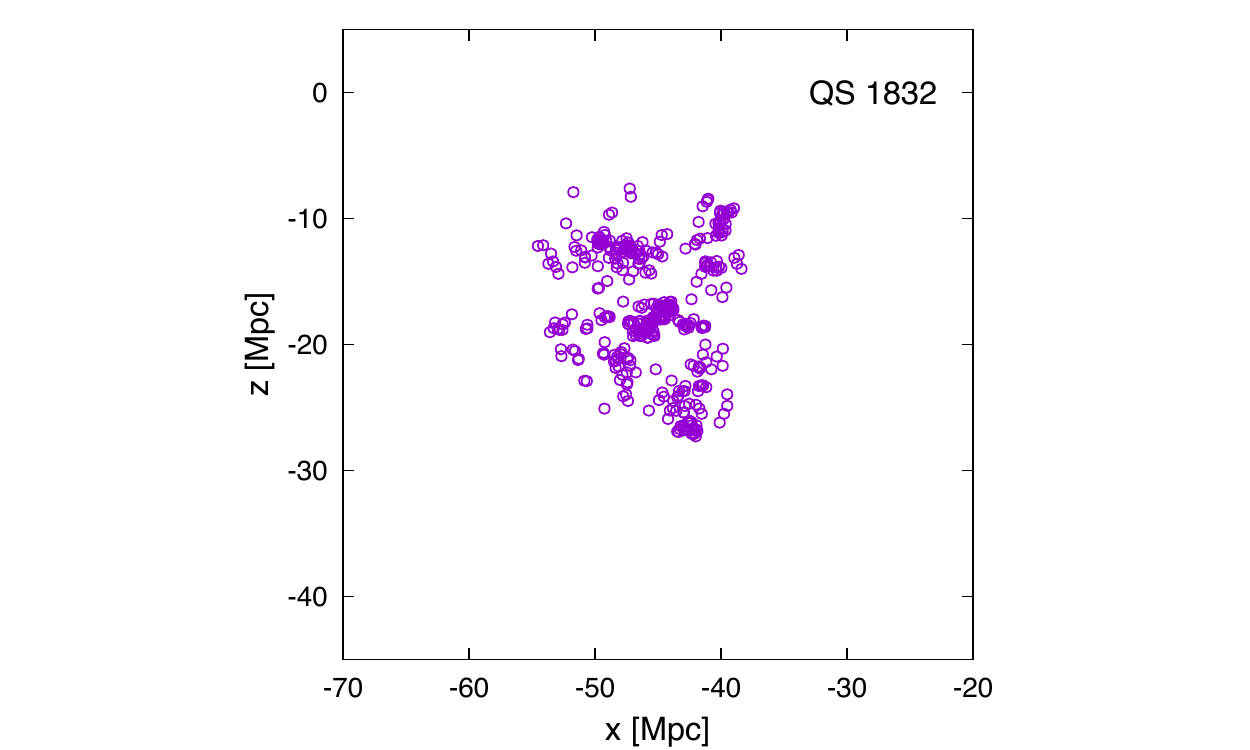}} 
\end{minipage}
\begin{minipage}{0.3\textwidth}
\resizebox{7.0cm}{!}{\includegraphics[width = 1.0in,angle=0]{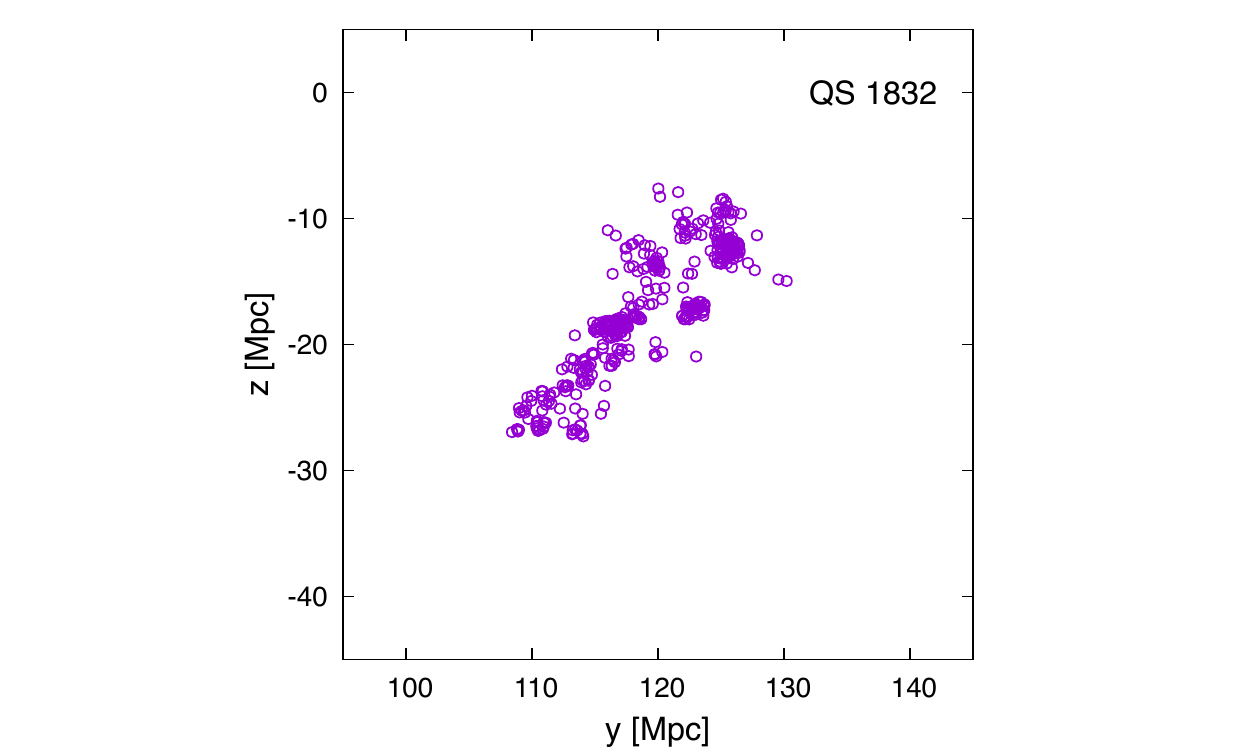}}
\end{minipage}\\
\caption {Distribution of galaxies in the regions of superclusters QS 550, QS 849, QS 524, and QS 1832 viewed from three different orientations (x-y, x-z, y-z) in Cartesian coordinates. Distribution of galaxies in the plane of the sky are shown in Appendix B.}
\label{Figmap1}
\end{figure*}

The result of the position-based cross-matching between QSs and known Abell clusters is shown in Table~\ref{table1}.
Figure ~\ref{Figmap1} shows the sky distribution of the galaxies  
in the region of the QS superclusters viewed from three different orthogonal directions in Cartesian coordinates. Before the transformation of the angular coordinates to the Cartesian one, redshift distortions in the groups and clusters were corrected as described in section 2. 
   
All QS superclusters contain a known Abell cluster in their central region. The cross-matching with other supercluster catalogues based on a fixed density level
showed that QSs can also be identified as a part of the previously known superclusters. 
QS 550 is in the Hercules supercluster or in the SCl 154 based on the catalogue of superclusters of Abell clusters by Einasto et al. (\cite{einasto2001}). 
QS 849 is in the zone of the supercluster SCl 541, extracted by a fixed density threshold limit by Liivam\"agi et al. (\cite{liivamagi12}).
QS 524 is located and centred on the richest cluster A1436 in the Ursa Major 
supercluster (Einasto et al. \cite{einasto2001}) and lastly, QS 1832 is in the Bootes
supercluster.

Clearly, the adaptive threshold density method extracts more concentrated superclusters, 
which represent high-density 
regions (or cores) of the superclusters defined by a fixed density level. The QSs have their masses in the range of $M \approx 1-3\times~10^{15} \,M_\odot$. This is the same mass range as Einasto et al. (\cite{einasto16}) found for the  components  (high-density cores) in the four superclusters of the Sloan Great Wall (SGW). These latter authors concluded that the core regions of some of these elongated components with radii smaller than $\approx$ 12 Mpc may already be collapsing.

\begin{table*}
\caption{Quasi-spherical superclusters: Coordinates and sizes}    
\label{table1}      
\centering          
\begin{tabular}{c c c c c c c}     
\hline\hline       
ID & $ID_A$ & R.A. & \it{Dec.} & z & Diameter (Mpc) & (Degrees)  \\ 
\hline                    
   550  & A2052 & 228.872 & 7.388  & 0.045414 & 32.9 & 9.8 \\  
   849  & A1983 & 233.348 & 17.387 & 0.045906 & 32.0 & 9.4  \\  
   524  & A1436 & 179.611 & 56.119 & 0.063341 & 24.7 & 5.4  \\
   1832 & A1797 & 207.739 & 23.528 & 0.031429 & 26.2 & 11.1  \\
  
\hline                  
\end{tabular}
\tablefoot{                                                                                 
Columns are as follows:
(1) ID of the supercluster in the catalogue; 
(2) ID of the main Abell cluster in the given QS;
(3)--(4) right ascension and declination of the centre of density in degrees;
(5) redshift of the marker galaxy: the brightest galaxy near the highest density peak in the supercluster;
(6) the maximum distance between galaxies in Mpc;
(7) diameter (in degrees)
}
\end{table*}

\begin{table*}
\caption{Quasi-spherical superclusters: Properties  }    
\label{table2}      
\centering          
\begin{tabular}{c c c c c c c c c}     

\hline\hline       
ID & $N_{groups}$ & $N_{galaxies}$ & $L_{tot}$($10^{10}L_\sun$) & $M_{tot}$($10^{15}M_\sun$) & $M/L$ ($M_{\sun}/L_{\sun}$) & $log(\langle\rho_{\mathrm{M}}\rangle / \rho_{\Lambda})$   \\ 
\hline                    
   550  & 50 & 560 & 680 & 2.7 $\pm$ 0.76 & 405 $\pm$ 112 & 0.22  $\pm$ 0.18\\  
   849  & 59 & 531 & 630 & 1.6 $\pm$ 0.32 & 262 $\pm$ 51\   & 0.034 $\pm$ 0.20 \\  
   524  & 29 & 283 & 440 & 1.9 $\pm$ 0.67 & 436 $\pm$ 154 & 0.43  $\pm$ 0.16 \\
   1832 & 54 & 422 & 330 & 1.2 $\pm$ 0.23 & 376 $\pm$ 69   & 0.17  $\pm$ 0.15 \\ 
   
\hline                  
\end{tabular}
\tablefoot{                                                                                 
Columns are as follows:
(1) ID of the supercluster in the catalogue; 
(2) number of groups;
(3) number of galaxies;
(5) total r-band luminosity;
(6) estimated dynamical mass of the system;
(7) estimated mass-to-light ratio;
(8) ratio of the average mass density of the system over the dark energy density
}
\end{table*}

An interesting point is that, contrary to the general trend of the properties, the $M/L$ ratios of the QS superclusters are not the 
smallest ones among the superclusters. Their mean $M/L$ ratio is about 
$364 \,M_{\sun}/L_{\sun}$, while for the rest of the superclusters of our sample
(white crosses in figure~\ref{FigSCC}) $\langle M/L \rangle =  319 \,M_{\sun}/L_{\sun}$. 
The $M/L$ ratios of the QSs are given in Table~\ref{table2}. 
We see that QS 849 has the lowest $M/L$ ratio. 
In Sect. 5, we show that QS 849 may also be in a different dynamical phase
from the other three QSs.  
If we leave out QS 849 and consider only QS 550, QS 524, and QS 1832, their mean $\langle M/L \rangle$  
is equal to $406 \,M_{\sun}/L_{\sun}$. This relatively high ratio may indicate an exceptional amount of dark matter in these systems.

In Table~\ref{table2}, we also present the number of galaxy groups and individual galaxies in each QS. The third 
column shows the estimated total mass of the QS. The mass errors are calculated as explained in Sect. 2.1.
We note that the most massive groups ($M_{tot} \geq 10^{14} \,M_{\sun}$)  dominate 
the total mass budget of the QSs, except in QS 849. 
The fraction of groups more massive 
than $10^{14} \,M_{\sun}$ is $77\%$ in QS 550, $72\%$ in QS 524, and $55\%$ in QS 1832. In QS 849, the majority ($63\%$) of its total mass is in groups with masses of less than $10^{14} \,M_{\sun}$.

\vspace*{0.0cm}

\section{The $\Lambda$ significance diagram}

Due to the large variation in supercluster morphology, a complex ensemble of substructures, and our almost total lack of information on the velocity fields outside of the local Universe, the study of the dynamical state of superclusters as a whole is difficult.
An exception to this, are superclusters that are relatively spherical and do not contain several massive substructures, similar to the four quasi-spherical superclusters found above.  
In those cases, an analytical approach like the spherical collapse model (e.g. Gramann et al. \cite{gramann15}; Einasto et al. \cite{einasto20}) or the $\Lambda$ significance diagram (Teerikorpi et al. \cite{teerikorpi15}) can be applied. Below, we use the $\Lambda$ significance diagram to estimate their dynamical state.

The dynamical state of the gravitating system in the expanding Universe can be analysed 
studying the motion of a test particle in the gravity field of the system treated as a point-like
mass (or a spherical mass distribution) in the background of the anti-gravitating force due to DE.
The main parameter in this approach is the
zero-acceleration radius $R_{\rm ZG}$
(e.g. Chernin \cite{chernin01}; Chernin et al. \cite{chernin06}):

\be
 R_{\rm ZG} = (\frac{3GM}{\Lambda c^2})^{1/3} =
(\frac{3M}{8\pi \rho_{\Lambda}})^{1/3},  
\ee

\noindent
which refers to the radius (in real space in 3D) where the radial peculiar component of the test particle velocity is equal to the Hubble expansion velocity in the gravitation field of mass $M$.  
A useful parameter, which measures the influence of DE, is the energy density ratio
$\langle \rho_{\mathrm{M}} \rangle / \rho_{\Lambda}$, as calculated for
the system under inspection. For that purpose, Teerikorpi et al. (\cite{teerikorpi15})
introduced the $\Lambda$ significance diagram (a $log(\langle \rho_{\mathrm{M}} \rangle / \rho_{\Lambda}) $ versus$\log R$ graph) where  $R$ is the
radius of a system, $\langle \rho_{\mathrm{M}}\rangle$ is its average mass density, and
$\rho_{\Lambda}$ is the DE density equal to the global value $\rho_{\Lambda} \approx 6 \times 10^{-30}$ g cm$^{-3}$.

The location of a galaxy system (or an inner part of it) in the 
$\Lambda$ significance diagram
indicates whether its overall dynamics is dominated by gravity or by the outward
`antigravity' expulsion of DE.
The diagram also displays a few relevant scales, which appear for each fixed mass of a spherical or a slightly flattened system in the $\Lambda$CDM cosmology. In Teerikorpi et al. (\cite{teerikorpi15}),
a few example galaxy systems were also shown in
a similar diagram.
In the present paper, we also use this graph as a tool to characterise the significance of the DM gravity--DE expulsion interplay for the dynamics of quasi-spherical superclusters.

\begin{figure}[htbp] 
\centering

\includegraphics[height=7.5cm,angle=0,width=100 mm]{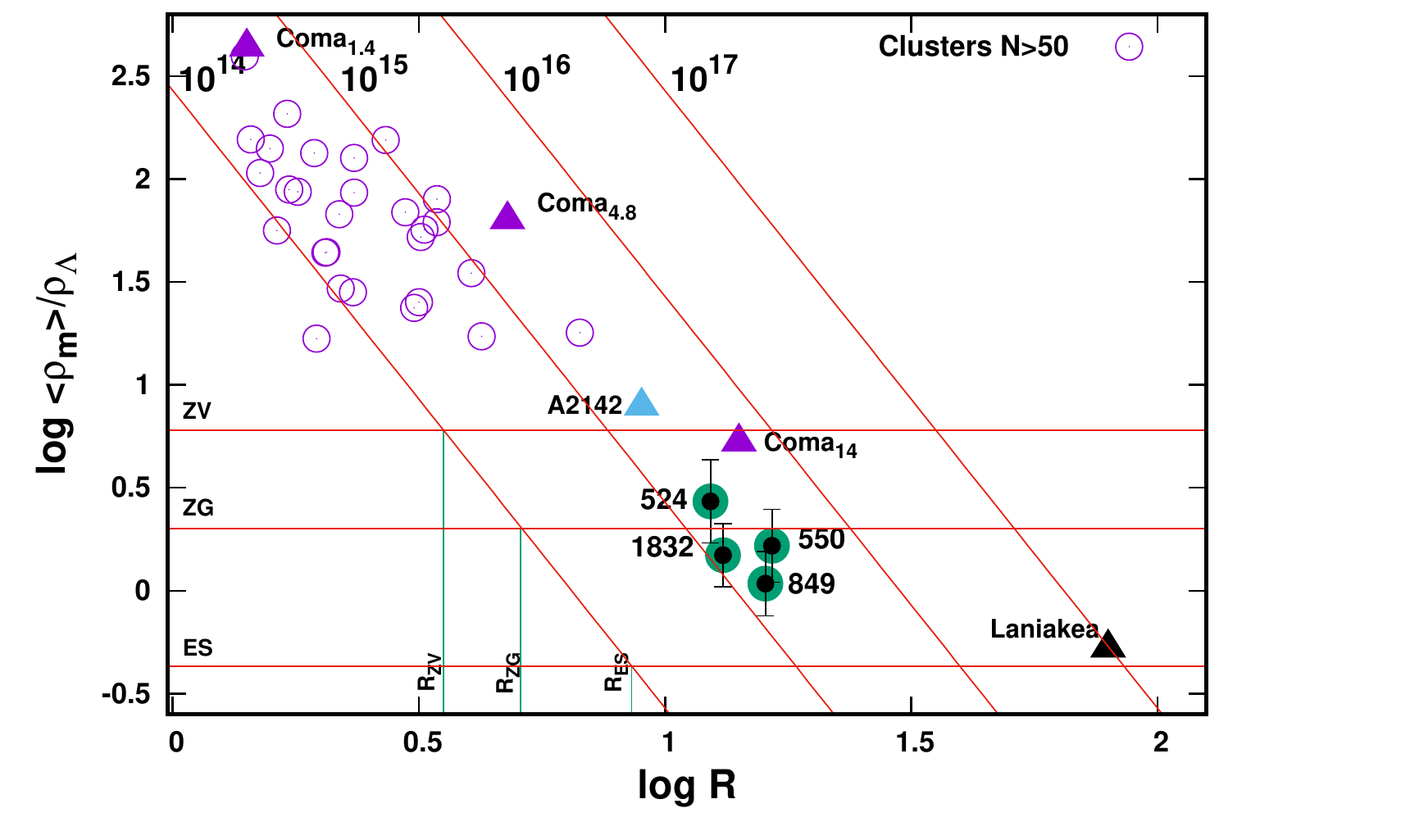}

\caption{ $\Lambda$ significance diagram for SDSS clusters of galaxies (purple) with QS superclusters marked in green. The Laniakea supercluster, the central regions of the supercluster A2142 and the Coma cluster with three different mass and size estimates are shown. Inclined lines show the effect of different example mass values. Green vertical lines show the values of the $R_{\rm ZV}$  (zero velocity radius), $R_{\rm ZG}$ (zero gravity radius), and the $R_{\rm ES}$ (the Einstein-Straus radius for the mass {$10^{14} \,M_{\sun}$}) and the horizontal lines show logarithmic values for the constant ratios of the $\langle \rho_{\mathrm{M}}\rangle / \rho_{\Lambda}$ = 6 ($\rm ZV$), 2 ($\rm ZG$), and 3/7 ($\rm ES$).}

\label{FigLamda1}
\end{figure}

In logarithms, 
the ratio $\langle \rho_{\mathrm{M}}\rangle / \rho_{\Lambda}$ becomes

\be
\log (\langle \rho_{\mathrm{M}}\rangle / \rho_{\Lambda}) = 0.43 + \log (M/10^{12}M_\odot)
- 3 \times \log (R/\mathrm{Mpc}),
\ee
where $M$ is the mass within the radius $R$ of a system.
In the $log(\langle \rho_{\mathrm{M}}\rangle / \rho_{\Lambda})$ versus $\log R$ diagram,
this relation forms a family of inclined straight lines for different values of 
the mass $M$.  The intersections of the lines with horizontal lines of constant
$log(\langle \rho_{\mathrm{M}}\rangle / \rho_{\Lambda})$ give the following scales or radii for the 
mass $M$:
\begin{enumerate}
\item The zero-velocity radius $R_{\rm ZV}$ (at this distance, a test particle system-centric radial velocity $u=Hr-v_{pec}$ is $v_{pec}=Hr$ and $u=0$.) This corresponds to the turn-around radius in the spherical collapse model.
\item The zero-gravity radius $R_{\rm ZG}$ (gravity force equal to Einstein’s  antigravity force). Thus, acceleration is zero, $du/dt=0$,  which indicates the
 maximum 
radius of a gravitationally bound system at the present epoch. 
\item The Einstein-Straus radius $R_{\rm ES}$ (the radial velocity reaches the Hubble velocity, $u=Hr$). This last, longest distance corresponds to the spherical volume where the mass $M$ produces an average density that is equal to the cosmic global density. 
\end{enumerate}
In Fig. ~\ref{FigLamda1}, these three different radii are shown for the systems of mass $10^{14} \,M_{\sun}$.
For spherical and homogenous systems, the horizontal lines ${\rm ZV}$, 
 ${\rm ZG,}$ and ${\rm ES}$ correspond to the 
constant ratios of $\langle \rho_{\mathrm{M}}\rangle / \rho_{\Lambda}$ = 6, 2, and 3/7. 

Figure~\ref{FigLamda1} shows different types of structures in the $\Lambda$ significance diagram. 
Clusters of galaxies (from the SDSS survey, closer than 450 Mpc) are shown with purple circles. These are well inside the $R_{\rm ZV}$, the turn-around radius in the spherical collapse model. Gramann et al. (\cite {gramann15}) and Einasto et al. (\cite{einasto15}) showed in their dynamical analysis that the supercluster A2142 has a relatively spherical core region (inside the radius of about 8 Mpc) and that it is in a collapsing state.  
We show the core region of the supercluster A2142 as a blue triangle. It is, as expected, close to but above the $\langle \rho_{\mathrm{M}}\rangle / \rho_{\Lambda}$ = 6 ($\rm ZV$) line and thus inside its zero-velocity radius $R_{\rm ZV}$.  According to Tully et al. (\cite{tully14}), after removal of the mean cosmic expansion and long-range flows, the Laniakea supercluster has a coherent, approximately spherical peculiar inward velocity flow field of matter (defining the basin of attractor) within a radius of about 80 Mpc. Outside this region, matter moves along the Hubble flow or is once again gravitationally attracted by some mass concentrations. Consistently, in Fig.~\ref{FigLamda1}, the Laniakea system  is located near the $\langle \rho_{\mathrm{M}}\rangle / \rho_{\Lambda}$ = 3/7 ($\rm ES$) line. 
The Coma cluster is known for its spherical symmetry.  Following Teerikorpi et al. (2015) and Chernin et al. (2013),  three different 
mass estimates  at different scales around Coma are plotted in Fig.~\ref{FigLamda1}. The Coma cluster seems to be inside the turnaround radius up to nearly 14 Mpc. 

The quasi-spherical superclusters found in this paper are marked with green dots and ID numbers. 
Only QS 524 is located inside the $R_{\rm ZG}$ radius and perhaps also QS 550, if the error bars are considered. 
The error bars for the $\log(\langle \rho_{\mathrm{M}} \rangle / \rho_{\Lambda})$ values are obtained using the estimated errors for masses in Sect. 2.1. 
The Finger-of-God effect is  during the group reconstruction procedure. We take into account the possible  
Kaiser effect including a 10\% error for the QS radii. We also note that the adopted radii are defined as half of the diameter (defined as the maximum distance between galaxies in a QS). Thus, any deviation from the spherical shape of the QS will decrease the actual radii and therefore the value of $\log (\langle \rho_{\mathrm{M}} \rangle / \rho_{\Lambda})$.  
Table~\ref{table2} shows the estimated $\log (\langle \rho_{\mathrm{M}} \rangle / \rho_{\Lambda})$ for all QSs. 

In view of the observations that suggest that the more local Hubble constant $H_0$ is somewhat larger than the Planck result (the `Hubble tension'), we note that the parameter $<\rho_M> / \rho_{\Lambda}$ (Eq. 3) has the useful property that it does not directly depend on $H_0$. This may be seen as follows. The dynamical (virial) mass for each subsystem of a supercluster is proportional to a size parameter, and therefore to $H_0 ^{-1}$, and then the density of the whole system is $\propto H_0 ^{-1} / H_0 ^{-3} = H_0 ^2$. On the other hand, the critical density is $\propto H_0 ^2$. Hence, in the ratio $<\rho_M> / \rho_{\Lambda}$ the strong quadratic dependence on $H_0 ^2$ cancels out and the only cosmological parameter that remains is $\Omega_{\Lambda}$ in the nominator of the ratio. We assume that $\Omega_{\Lambda}$ does not differ very much from its standard value. If it did, for example by 10 percent, then the shift in the y-axis of the $\Lambda$ significance diagram (Fig.~\ref{FigLamda1}) would be only 0.04. A different $H_0$ would make a shift along the x-axis, but being constant this does not affect our discussion.

\subsection{Shapes}

In the adaptive density threshold method, the superclusters 
were identified  using a 8 $h^{-1}$ Mpc kernel. Once the superclusters 
were extracted, the Minkowski functionals were applied to the smoothed (with a 4 $h^{-1}$ Mpc kernel) luminosity density field of the galaxies in each supercluster. This approach  efficiently sorts the general morphologies from the large variety of supercluster shapes. Smoothing also averages the shapes. Although we used a tight criterion for the shape-finder vectors $K_{1,2}$ to select the most spherical superclusters, all four QSs were found to be flattened at least in one axial direction  when only galaxy positions were used to determine the ellipticity. 
Therefore, we calculated the covariance matrix and eigenvalues for the galaxy distribution in each supercluster in order to define the principal semi-axes of the ellipsoid.

The superclusters SC 550 and SC 849 are found to be considerably elongated systems (namely $c > a$, $c > b$ and their $\beta = \frac{b}{a} = 0.84$), and may be interpreted as prolate spheroids, although the $\beta$ value indicates a mild triaxiality. 
The superclusters SC 524 and SC 1832 more closely resemble triaxial-type ellipsoids as they have  $\beta =\frac{b}{a} = 0.61$ and $0.53$, respectively.

We note that 3D ellipse fitting to the galaxy positions results in more precise shapes than the method based on the combination of the luminosity density and the Minkowski functionals. 
In the former method, 3D ellipse fitting, galaxies are  considered as points, while the latter method instead follows the shape of the galaxy luminosity field, which represents the mass and therefore the gravitational potential field of the supercluster. Presumably, the real shape of the gravitational potential field of a QS supercluster is somewhere between the results given by these two different methods. We also tested the convex hull method; it gives more spherical shapes for all QSs.

\subsection{Zero-gravity radius for non-spherical spheroids}

\begin{figure}[htbp] 
\centering

\includegraphics[height=7cm,angle=0,width=100 mm]{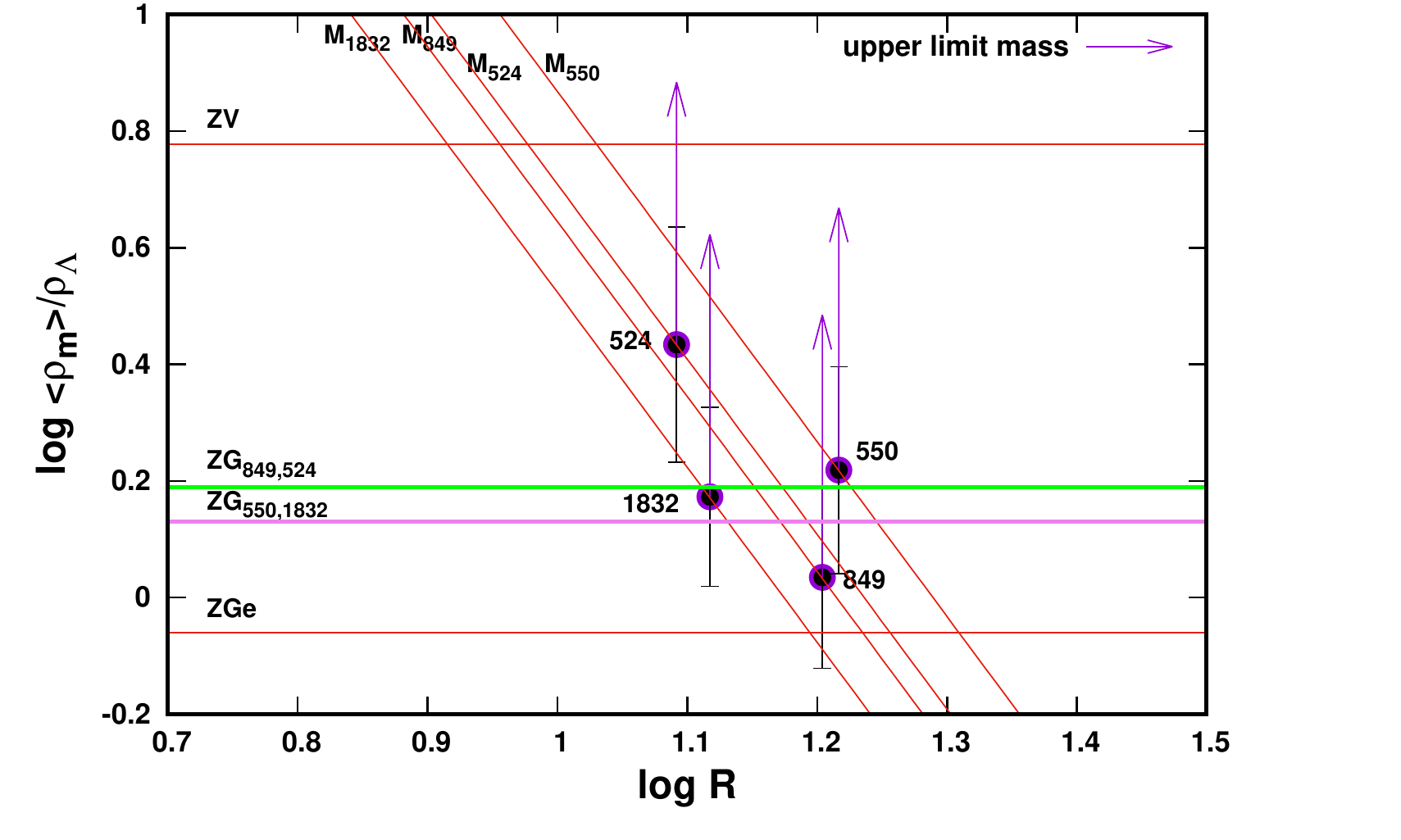}

\caption{ $\Lambda$ significance diagram for four QS superclusters.
The purple horizontal line shows the shape-corrected $\rm ZG$ balance line for QS 550 and QS 1832 and the green line shows the $\rm ZG$ value for QS 849 and QS 524. Inclined lines show the QS masses (see table~\ref{table2}).  
 Arrows show estimated upper limits for the QSs in the graph.
The  $\log (\langle \rho_{\mathrm{M}} \rangle / \rho_{\Lambda})$ $\rm ZV$ line is shown as in Fig.~\ref{FigLamda1}. The ${\rm ZGe}$ shows the theoretical limit of the zero-gravity balance line for completely flat systems.}
 
\label{FigLamda2}
\end{figure}

In Fig.~\ref{FigLamda1}, we assume that the QS superclusters are spherical. However, our analysis of the shapes (Sect. 5.1) shows that the QSs are relatively prolate or triaxial spheroids.  We know that   
for a flattened (oblate) system, the zero-gravity balance line  $log(\langle \rho_{\mathrm{M}}\rangle / \rho_{\Lambda})$ = 0.3 ($\rm ZG$) will be shifted down. In the limiting case of a homogenous infinitesimally flat disc,  the balance line $\rm ZG$ should be shifted down by 0.37 so that $log(\langle \rho_{\mathrm{M}} \rangle / \rho_{\Lambda})= -0.07$ (Teerikorpi et al. 2015). In Fig.~\ref{FigLamda2}, this line is marked with the symbol $\rm ZG_e$. 
Not only the amount of flatness, but also the matter distribution inside a non-spherical system affects the zero-gravity balance line. In practise, this means that all systems have their own individual zero-gravity balance line and thus the zero-gravity radius. 
A more detailed mathematical description of how these properties shift the position of the balance line $\rm ZG$ in the $\Lambda$ significance diagram is given in the Appendix. Here we 
give a brief summary. 

At first, we calculate the force factor $f$, which denotes the force ratio between 
a spheroid (oblate or prolate) and an ideal spherical system with the same mass and radius (along the longest axis of the system). This force factor is calculated for 
different combinations of axial ratios and density profiles. 
For oblate systems, the force factor $f(e,n)$ can be described as a function of the flatness (defined as $e^2= 1-(b/a)^2, a=c$) and the matter density profile $\rho=w^n$ of the system.
This results in the  balance line  $\Delta \langle \rho_{\mathrm{M}} \rangle / \rho_{\Lambda}$ which shifts smoothly as a function of the flatness. The shift is 
larger for systems with a flatter shape and flatter density profile (highlighted in Fig.~\ref{FigA2}).

Next we estimate how the  balance line $\rm ZG$ will be shifted for prolate systems. 
First, we consider the force factor $f_{\alpha}$ for prolate homogenous spheroids with different axial ratios ($\alpha=0$ denotes homogenous mass distribution).
We also consider different cases where a fraction of the total mass of a system is located at its centre and the rest is distributed uniformly. Presumably, the real distribution is somewhere between the cases $\alpha=0$  and   $\alpha=0.5$ (half of the mass at the centre; see Appendix A.2 and Table A.1).

In Fig.~\ref{FigLamda2} we show how the balance line $\rm ZG$ of each QS is shifted when $\alpha=0.5$ is assumed. In such a case, only QS 849 is clearly below its balance line $\rm ZG$ (green horizontal line) and therefore it is also outside of its $R_{\mathrm{ZG}}$. Three other QSs are above their individual $\rm ZG$ line. Our analysis of the  axial ratios in Sect. 5 showed that  
QS 524 and QS 1832 are relatively triaxial while QS 550 is quite a prolate system. 
Thus, QS 524 and QS 1832 are 
not pure prolate or oblate systems, but between these two types of spheroids.
Figure~\ref{FigA2} shows the general trend that the shift $\Delta \langle \rho_{\mathrm{M}} \rangle / \rho_{\Lambda}$ increases when moving from spherical to oblate systems and further to prolate ones. 
In Fig. ~\ref{FigLamda1}, the QSs are considered as purely spherical systems and those in Figure~\ref{FigLamda2} as purely prolate systems. This indicates that the actual balance lines $\rm ZG$ for QS 524 and QS 1832 are situated between these two extremes. 
In that case,  QS 1832 might not be inside its zero-gravity radius $R_{\mathrm{ZG}}$.\\
It is also important to note that the virial masses of the component galaxy groups (and hence the supercluster masses) 
might be systematically underestimated. This is because 
the size and virial mass of a system will be undervalued if the dark
halo is not fully probed by the observed galaxies (Chernin et al. \cite{chernin12}). The possible
amount of mass missed in this way was studied by Teerikorpi et al. (\cite{teeri18}) who concluded that,
in comparison with the mass up to $R_{200}$, the maximum mass could be roughly  a factor of two larger.
Karachentsev et al. (2017) considered the missing halo mass around galaxy groups, including also
the possible retarded outflow component up to the Einstein-Straus distance. These authors `calibrated' this fraction
using the mass and average matter density determinations in the local Universe, as compared with
the average global dark matter density. The conclusion was that the total mass of a typical group, including
its extended halo, could be a factor of three larger than the usually determined virial mass.  
This is shown in Fig.~\ref{FigLamda2} with arrows pointing to the positions in the $\Lambda$ significance diagram when the supercluster 
masses are multiplied by a factor of three, corresponding to a plausible upper limit. 
As mentioned before, previous studies have shown that the group mass estimation method adopted by us also systematically underestimates masses. 
If the masses of the QS superclusters were for example two times larger than adopted here, then they 
would be found close to or above the $\rm ZV$ limit line 
for spherical systems. This corresponds to the turnaround radius, the boundary where the system completely 
decouples from the dark-energy-driven expansion and virialized. 
All this means
that $\log (\langle \rho_{\mathrm{M}} \rangle / \rho_{\Lambda})$ values in Fig.~\ref{FigLamda2} can be 
considered as lower limits and it is probable that 
at least QS 524 and QS 550 are gravitationally bound objects. 
From a dynamical point of view, these objects represent intermediate systems between clusters and superclusters in general.\\
While zero gravity puts a minimum limit on the overdensity of a gravitationally bound system 
with mass and size in the current epoch, it does not necessarily prevent the system from expanding 
in the future. The futures of collapsing systems in the simulations and observations have been studied 
previously, by  for example D\"{u}nner et al. (\cite{dynner06}), Luparello et al. (\cite{luparello11}), Chon et al. 
(\cite{chon15}), and Gramann et al. (\cite {gramann15}). In these studies, the authors showed that, to ensure 
the collapse of the system in the future, the characteristic overdensity of the systems at the present 
epoch ($\Omega_m=0.3$ and $\Omega_{\Lambda}=0.7$) should be at 
least $<\rho>/\rho_m \approx 7.8$ or $<\rho>/\rho_c \approx 2.4$ or $<\rho>/\rho_{\Lambda} \approx 3.4$. Simulations show that in the future, such 
systems will get more and more spherical and finally collapse. The value of the density contrast for the future collapsing system at the present epoch $\log (<\rho>/\rho_{\Lambda}) \approx 0.5$ is somewhat higher than the values found 
for QS 524 $\log (<\rho>/\rho_{\Lambda})=0.43$ (see Table~\ref{table2} and Fig.~\ref{FigLamda2}). This means that we cannot predict whether or not QS 524 will collapse in the future.

\section{Discussion}
\subsection{Gas in superclusters}

The warm-hot intergalactic medium (WHIM) is 
expected to be distributed in relatively low-density non-virialized stuctures within
superclusters and filaments of galaxies (Cen \& Ostriker \cite{Ceno}). In principle, the all-sky $y$-maps of the diffuse gas can be used to trace WHIM. However,  $y$-values of at least two orders lower as compared with those of clusters decrease the signal to the noise level, hampering the analysis. \\
To avoid this problem,    
Tanimura et al. (2018) resorted to statistical analysis and stacked the Planck $y$-maps of the regions of  580 superclusters.
After careful masking of the clusters and groups they reported the first detections of the intercluster gas in superclusters at the 2.1$\sigma$ level. 
In general, superclusters have very multimodal shapes, which 
unfortunately increases the noise.     
We realised that this effect is minimised for QS superclusters.
Secondly,  we found above that the QS superclusters are in or near to the ZG radius and thus at the gravitationally bound stage.
For such systems, the density contrast
 (the mean matter density of the system over the mean matter density of the Universe) $<\rho>/\rho_{m}\approx 5$ or $<\rho>/\rho_{c}\approx 1.5$ is required (Gramann et al. \cite {gramann15}). 
According to simulations, this value approximately corresponds to the baryon 
overdensity $<\rho/\rho_{b}>\approx 10$ (Cen \& Ostriker \cite {cen06}) or $<\rho/\rho_{c}>\approx 0.2$ (Haider et al. \cite {haider16}) of the system and this in turn corresponds to the lower limit of the baryon density where most of the WHIM mass between temperatures $10^{5}-10^{7} \,K$ should be located.
Therefore, it would be interesting to see whether or not the QSs 
give a positive Sunyaev-Zeldovich signal.

With this aim in mind, we follow the procedure used by Tanimura et al. (2018): 
First, we stack the four supercluster positions in the Planck y-map (Aghanim et al. \cite{planckymap}) in taking into account the cluster mask, the point source mask, and the ICB mask of Planck. We then compute the aperture photometry of the inner part. Finally, 2000 random stacks are made to build the probability distribution of the photometry.

\begin{figure}[htbp] 
\centering

\includegraphics[height=7cm,angle=0]{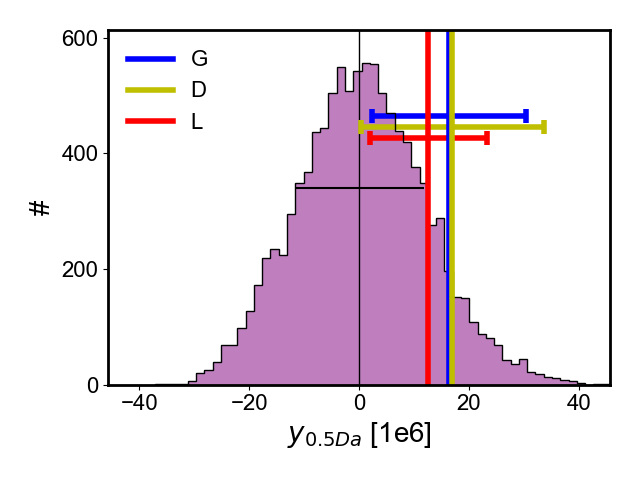}

\caption{y-signal distribution of the random sample superimposed by the four stacked QS superclusters. G: location of the galaxy near the highest density peak in the supercluster, D: position of the density peak of the supercluster, L: luminosity centre of the supercluster.}
\label{FigSZ}
\end{figure}

Figure~\ref{FigSZ} shows the result for the four QSs using three different centre
definitions compared with random positions.
The horizontal error bars
are obtained by bootstrap. The photometry is taken at half of the diameter (see Table~\ref{table1}); 
this was found to give the most prominent signal.
According to our analysis (Fig.~\ref{FigLamda2}),   
the radius R of the QSs approximately corresponds to the zero-gravity radius of these systems. 
Their zero-velocity radius is about half of the $\rm ZG$ radius. 
Figure~\ref{FigSZ} shows that the errors are large and the signal is marginal (about 1.5$\sigma$ detection).
However, even this result is promising, because we have used only four objects for stacking.

\subsection{Dark matter in superclusters} 

The mass-to-light ratios determined for clusters and superclusters show large variations, from about 150 to several hundred (e.g. Bahcall et al. \cite{bahcall95}; Carlberg et al. \cite{carlberg}; Gvazzi et al. \cite{gavazzi}; Schirmer et al. \cite{schirmer}; Bachall \& Kulier \cite{bahcall14}; Einasto et al. \cite{einasto15}, \cite{einasto16}). Despite the scatter, the results imply that the mean mass-to-light ratio $\langle M/L\rangle_{r}$ increases as a function of scale, 
and even at cluster scales begins to approach the Universal value (Bachall \& Kulier \cite{bahcall14}).
The fact that $\langle M/L\rangle$  remains about the same from clusters to superclusters
indicates that the superclusters do not contain a significant amount of additional dark matter
between the member clusters (we considered this point  when estimating
supercluster masses from the summed dynamical masses of the member systems).
According to Blanton et al. (\cite{blanton}), the SDSS DR7 mean luminosity density in the r-band is $\bar{L}_{r} \simeq 1.2 \pm 0.14 \times 10^8 \,L_{\sun}$ Mpc$^{-3}$. Using this value and with $\Omega_{\rm m} = 0.3$, $\Omega_{\Lambda}=0.68,$ 
the mean r-band mass-to-luminosity ratio on
cosmologically representative scales
will be $\langle M/L\rangle_{r} \simeq 330 \pm 40 \,(M_{\sun}/L_{\sun})$. This value is close to the average supercluster mass-to-light ratio $\langle M/L\rangle_{r} \simeq 319 \,(M_{\sun}/L_{\sun})$ found in Sect. 4. 
A considerably higher $\langle M/L\rangle_{r}$ value was found for the  
smallest and the most spherical superclusters in the sample. This may indicate that the QS superclusters contain more dark matter than superclusters in general, in analogy with dwarf galaxies on much smaller scales.

\section{Summary}

The elongated distribution of data points in the diagram of Fig.~\ref{FigLamda1} has, 
not unexpectedly, a similar  slope to that of the de Vaucouleurs (\cite{vauco70}) 
universal density--radius relation ($-1.7$), which he showed to extend 
from compact galaxies up to the Local supercluster and the local Universe 
as known at the time. A rather similar result was obtained by 
Karachentsev (\cite{kara68}) from pairs of galaxies up to superclusters.
This relation, showing that smaller systems have a higher mean density, evidently 
reflects the hierarchical structure of the galaxy Universe, suggesting that 
the groups appearing in the diagram, for example, are actually parts of larger systems and 
are therefore naturally denser than their host systems, on the average.

The SDSS groups as well as prominent clusters in general
are usually located deep within the 
gravity-dominated radius ($R \ll R_{\mathrm{ZV}}$), which is also a requirement
for them to be virialized. The virialized region starts when 
$log \langle \rho_{\mathrm{M}}\rangle / \rho_{\Lambda} > 0.78$.
The systems above the zero-velocity radius line are totally within
the collapsing region. The outer parts of the systems below the line have not yet
been retarded down to zero velocity. For example, Einasto et al. (\cite{einasto20}) estimated that the
$R_{\mathrm{ZG}}$ scale for the supercluster A2142 is about 11.5 Mpc.    

In reality, there could be gravitationally bound superclusters that are not spherical, but 
have some other type of morphology. 
In any case, it is interesting that the superclusters with almost spherical 
morphology appear to be  
gravity dominated. This is an extension of the general trend:
the gravitationally bound celestial bodies, from planets and stars to galaxies
and galaxy clusters, tend to exhibit spherical or near-spherical
symmetries
or axial-symmetries,
which may be traced backed to the isotropic effect of 
gravity around a mass concentration. This is also the case for 
systems still collapsing under gravity.

Dynamically, the large-scale structures of the Universe evolve under an interplay between two factors, dark energy and gravitational force.
We used a sample of 1313 superclusters constructed using the 3D luminosity density field and 
an adaptive threshold method (Liivam\"{a}gi et al. {\cite{liivamagi12}). This method makes it possible to determine for each supercluster  
its own density threshold limit. As the cosmological density field and peculiar velocity fields are closely related, we assume that the adaptive density threshold method may provide a way to  
extract superclusters that are physically more interconnected than systems found using a fixed 
density threshold method.
After consideration of various selection effects, we selected a subsample of 65 superclusters for our final analysis. 
Using the Minkowski functionals, we studied supercluster morphologies as a function of their physical 
properties, such as total mass, luminosity, and volume. 
We extracted a small sample of the most spherical (quasi-spherical) superclusters. 
The $\Lambda$ significance diagram was originally intended for spherical systems (similarly to the spherical collapse model). 
However, we also found simple analytical solutions for non-spherical spheroids in order to also  characterise their dynamical state using this diagram.  

The results of our analysis of superclusters extracted with an adaptive local threshold density method can be summarised as follows: 1) Superclusters with roughly spherical or pancake-type shapes are  smaller, less massive, less luminous, and contain less galaxies and groups than other superclusters. 2) In our sample of 65 superclusters, we found four potential QS superclusters indicating that such systems are relatively rare, representing just a few percent of all superclusters.   
These systems have relatively high mass-to-light ratios, 
which may indicate that these small and exceptionally spherical superclusters contain 
relatively large amounts of dark matter. 
The QSs found in this work are identified as 
high-density regions of the known extended superclusters defined with a fixed density level. 3) The thermal Sunyaev-Zeldovich effect signal found using the four stacked QS superclusters is not formally significant, but promising enough to encourage a follow-up study of selected superclusters and stacking of their $y$-signals. 
For example, the number of systems can be increased by allowing a wider range for the $K1$ and $K2$ parameters and the clumpiness of the system. 4) Using the $\Lambda$ significance diagram and lower limits for the masses of the QSs, we found that  
two of them, QS 524 and QS 550, and perhaps also QS 1832, are dynamically bound 
systems that form their own entity. If so, then these objects are among the largest gravitationally bound systems found to date. 

QSs are large, symmetrically bound systems with a smooth matter
distribution. This indicates special smooth initial conditions and
a `quiet' formation history. 
QSs represent a special class of giant systems that are dynamically between large
gravitationally unbound superclusters and clusters of galaxies in an equilibrium configuration. Comparing our   
 observational results with data from large cosmological
simulations could be an interesting cosmological test.

\section*{Acknowledgements}
We thank the referee for a report which helped us to improve the paper.
We thank Juhan Liivamägi for helpful discussions.
This work was supported by grants of the Vilho, Yrjö and Kalle Väisälä Foundation of the Finnish Academy of Science
and Letters. The present study was also supported by the ETAG project 
PRG1006,  and by the European Structural Funds
grant 79 for the Centre of Excellence "The Dark Side of the Universe" (TK133). We applied in this study the R statistical environment (Ihaka \& Gentleman 1996)

\begin{appendix}

\section{Zero-gravity radius for oblate and prolate matter spheroids}

The $\Lambda$ significance diagram is originally intended for spherically symmetric systems, but it also has limited use for flattened or elongated spheroids as we show here. 
A system cannot  be bound as a whole if the antigravity
force between its centre and a part of the edge is greater than the gravity
pulling the edge inwards. Here we consider the gravity force at the rim of a disc (an oblate spheroid) and at the end of a `cigar' (a prolate spheroid), in comparison with the corresponding antigravity force.

 It is expected that, at the rim of the disc and at the end of the cigar, the gravity force is larger than at the surface of a sphere of the same mass and the radius equal to the corresponding (half) major axis of the spheroid. This may be understood by noting that when a sphere is transformed into a spheroid, the average distance of the mass points from the considered fixed point on the surface of the sphere is decreased.  Also, the projections of the forces along the considered axis are larger. The change is bigger for a prolate spheroid.  

\subsection{Oblate spheroid}

A disc exerts a higher gravity force on a particle at its rim than a sphere  of the same mass and radius. If the ratio of these
forces is $f (>1)$, then  the balance condition at the equatorial rim $R = a$ is

\be
f \times \frac{GM}{a^2} = \frac{8\pi G}{3}
\rho_{\Lambda} a  
,\ee
or
\be
 \langle \rho_{\mathrm{M}} \rangle = \frac{2}{f} \rho_{\Lambda}
.\ee
 Here, $\langle \rho_{\mathrm{M}} \rangle $ is the mean density
caused by the mass $M$ within the sphere of radius $a$
($<$ the mean density of the disc). Formally, it may be used in the $\Lambda$ significance diagram, provided that
the `gravity = antigravity' line is correspondingly lowered for a flattened system, which here we take to be a spheroid with the axes $a=c$ and $ b < a$. We consider the case of homogenous mass distribution as well as the case of the density decreasing from the centre.

We calculate the factor $\alpha = f(e,n)$ for spheroids characterised by the eccentricity $e$ (defined as $e^2 = 1- (b/a)^2$) and the density decreasing from the centre as the power law $w^n$, where $-3 < n \leq 0$ and $w$ parametrizes the spheroidal shells inside the outer surface. The value $n=0$
corresponds to a homogenous spheroid. It is known (see e.g. Woltjer 1967; Hofmeister et al. 2018) that then the gravity force at the equator, at the edge of the spheroid with the equatorial radius $a$ is 

\be
F_{eq} = \frac{GM}{a^2}\frac{n+3}{e^{n+3}} \int_0^{\arcsin e} sin^{n+2}\theta d\theta 
,\ee
hence
\be
f(e,n) = \frac{n+3}{e^{n+3}} \int_0^{\arcsin e} sin^{n+2}\theta d\theta
.\ee
For different values of $n$ (here $0,-2$), this allows us to calculate the factor $f(e,n)$ and hence to see how sensitive the location of  the `gravity = antigravity' line is when the eccentricity changes from the value $= 0$ (a sphere) towards more flattened spheroids. These are: 
\be
f(e,0) = \frac{3/2}{e^3} (\arcsin e - e (1-e^2)^{1/2}),
\ee
%
%

\be
f(e,-2) = \frac{1}{e} \arcsin e.
\ee

For a flat disc, with the density increasing inwards, the ratio $f(1,n) < 3\pi/4$, and
the case of constant density gives $f(1,0) = 3\pi/4 = 2.35$. In this limiting case, the `gravity = antigravity'  balance line in the $\Lambda$ significance diagram would be shifted down to the value  $\langle \rho_{\mathrm{M}} \rangle / \rho_{\Lambda} = -0.07$
in order to roughly take very flattened systems into account. Figures A1 and A2 show results for less flattened spheroids as a function of the axis ratio $b/a = (1-e^2)^{1/2}$.

We note that when the zero-gravity  condition is fulfilled at the equatorial rim of the spheroid, this  guarantees that the whole spheroid lies in the gravity-dominated region, because at the polar point the gravity force is greater than at the rim and the distance to the centre is smaller.

\subsection{Prolate spheroid}

Analogously to our analysis of oblate spheroids, we want to calculate the ratio $f$ of the force at the end of an elongated spheroid (having the axes $a$ = $b$ and $c > a$) and the force at the surface of a sphere of the radius $c$, both having the same mass $M$. The density of the spheroid is then $M/((4/3) \pi a^{2}c$). 

MacMillan (1930) discussed a general case of the potential field around a prolate homogenous spheroid. Here we are satisfied with calculating the gravitation force at the end of such a spheroid (density $ = \rho$), which also serves to check the result by MacMillan in this special case.
Using the general potential function as derived by MacMillan (1930) in his equation 32.13, by derivation along the z-axis at $z = c$  we obtain the following formula  for the present
case:

\be
F = \frac{3M}{c^2 e^3} (\frac{1}{2} \ln \frac{1+e}{1-e} - e)
.\ee
%
Here we have in addition used the expression $M = (4/3)\pi a^2c)\rho$ for the mass of the homogenous spheroid. The eccentricity is  $e^2 = 1- (a/c)^2$

Using this expression, the excess factor $f$ for a homogenous prolate spheroid becomes simply
\be
f = \frac{3}{e^3} (\frac{1}{2} \ln \frac{1+e}{1-e} - e).
\ee
It may be confirmed that  $f \rightarrow 1$, when $e \rightarrow 0$ (a sphere). Also, we  checked the formula for $F$ by direct numerical integration, summing the forces from thin circular discs, making up the spheroid. The resulting total force is found to be the same.

If a more realistic mass distribution is such that a fraction $\alpha$ of the total mass $M$ is situated as a point mass at the centre and the remaining mass fills the spheroid uniformly, then the ratio $f_{\alpha}$ is equal to

\be
 f_{\alpha} = \alpha + (1-\alpha) f_0
,\ee
where $f_0$ comes from the previous equation.
Table A.1 shows the ratios $f_0$ and $f_{\alpha}$ ($\alpha = 0.5$) for different axis ratios $a/c$.  As expected, the excess factor $f$ is larger at the end point of an elongated spheroid than at the rim of a flattened spheroid of the same maximum radius. This is because in the former case the mass is closer, on average, to the surface point considered. 
Figures A1 and A2 show the results of our calculations as graphs.

\begin{figure}
\centering

\includegraphics[height=7cm,angle=0, width=9cm]{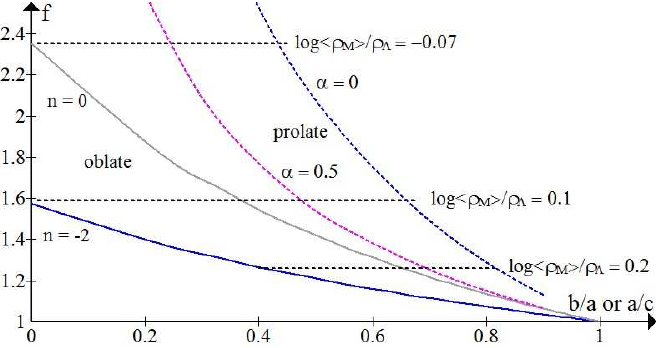}
\caption{Force excess factor $f$ vs. the axis ratios $b/a$ (oblate) and $a/c$ (prolate) of a spheroid for different density laws ($n=0$ and $\alpha = 0$ mean homogeneity). The corresponding values of $\log \langle \rho_{\mathrm{M}} \rangle / \rho_{\Lambda}$, when there is the `gravity = antigravity' balance at the maximum extension of the  spheroid,  are shown for three values of $f$. We note that for the spherical case ($b/a = a/c =1$) $f=1$, and $\log \langle \rho_{\mathrm{M}} \rangle / \rho_{\Lambda} = 0.3$, as generally used in the $\Lambda$ significance diagram. 
}
\label{FigA1}
\end{figure}

\begin{figure}
\centering

\includegraphics[height=7cm,angle=0,width=9cm]{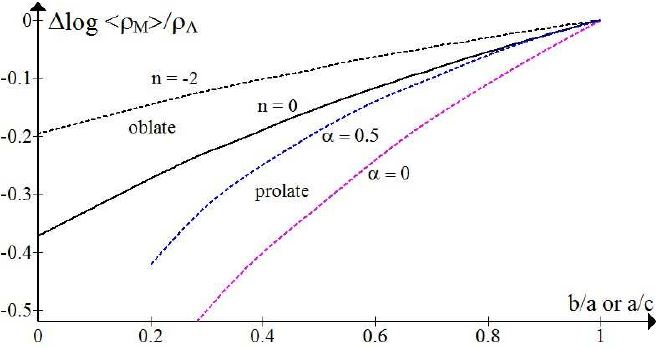}
\caption{Shift of the `gravity = antigravity' balance line in the $\Delta \log \langle \rho \rangle / \rho_{\Lambda} $ vs.
$\log R$ diagram as a function of the axis ratio $b/a$ (oblate) and $a/c$ (prolate). For example, when $a/c$ is greater than 0.5, the shift is less than about $0.3$ for all the cases considered.
}
\label{FigA2}
\end{figure}

\begin{table}
\caption{Excess force factors $f_0$ and $f_{0.5}$ for different $a/c$,}
\begin{center}
\begin{tabular}{ c c c }
\hline
$a/c$ & $f_0$ & $f_{0.5}$ \\
\hline
 1 & 1.0 & 1.0 \\ 
 0.9 & 1.13 & 1.07 \\  
 0.8 & 1.29 & 1. 15 \\
 0.7 & 1.49 & 1.25  \\
 0.6 & 1.75 & 1.38  \\
 0.5 & 2.08 & 1.54 \\
 0.4 &  2.53 & 1.77 \\
 0.3 & 3.18 & 2.09 \\
 0.2 & 4.20 & 2.60 \\
\hline
\end{tabular}
\end{center}
\label{table3}
\end{table}

\vspace*{0.0cm}\begin{figure*}[htbp]
\centering

\section{Quasi-sphericals in the plane of the sky}

 \begin{minipage}{0.45\textwidth}
\resizebox{8.0cm}{!}{\includegraphics[width = 1.6in,angle=0]{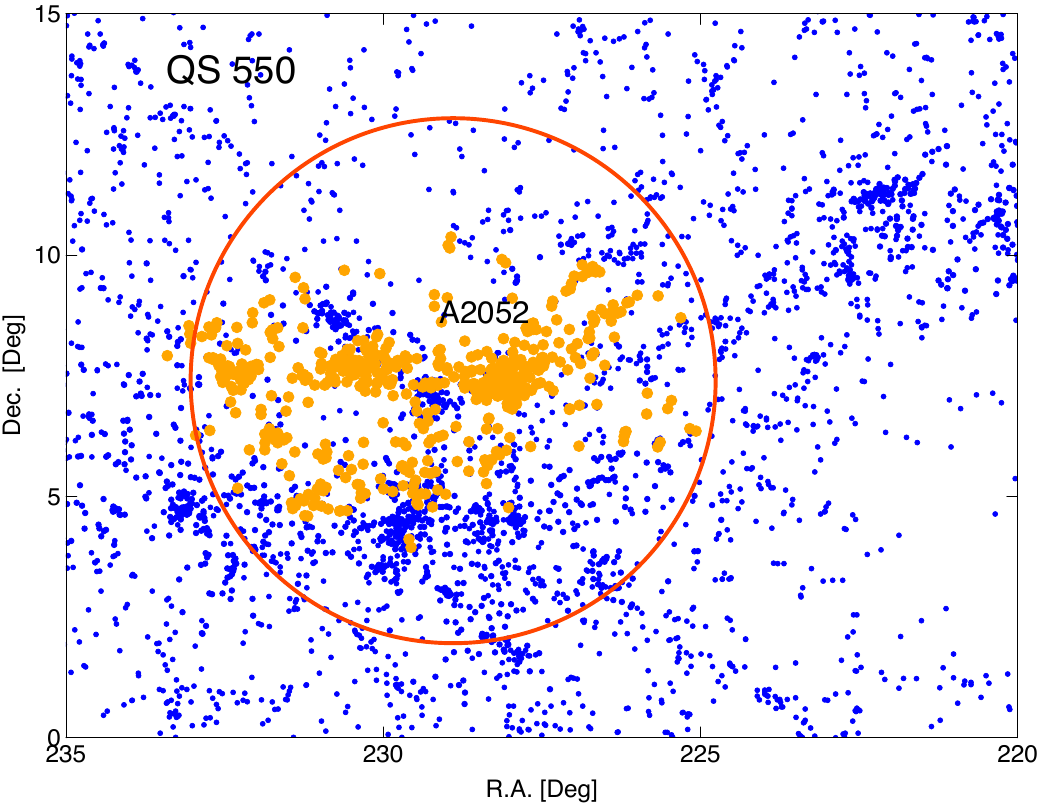}}
\end{minipage}
\begin{minipage}{0.45\textwidth}
\resizebox{8.0cm}{!}{\includegraphics[width = 1.6in,angle=0]{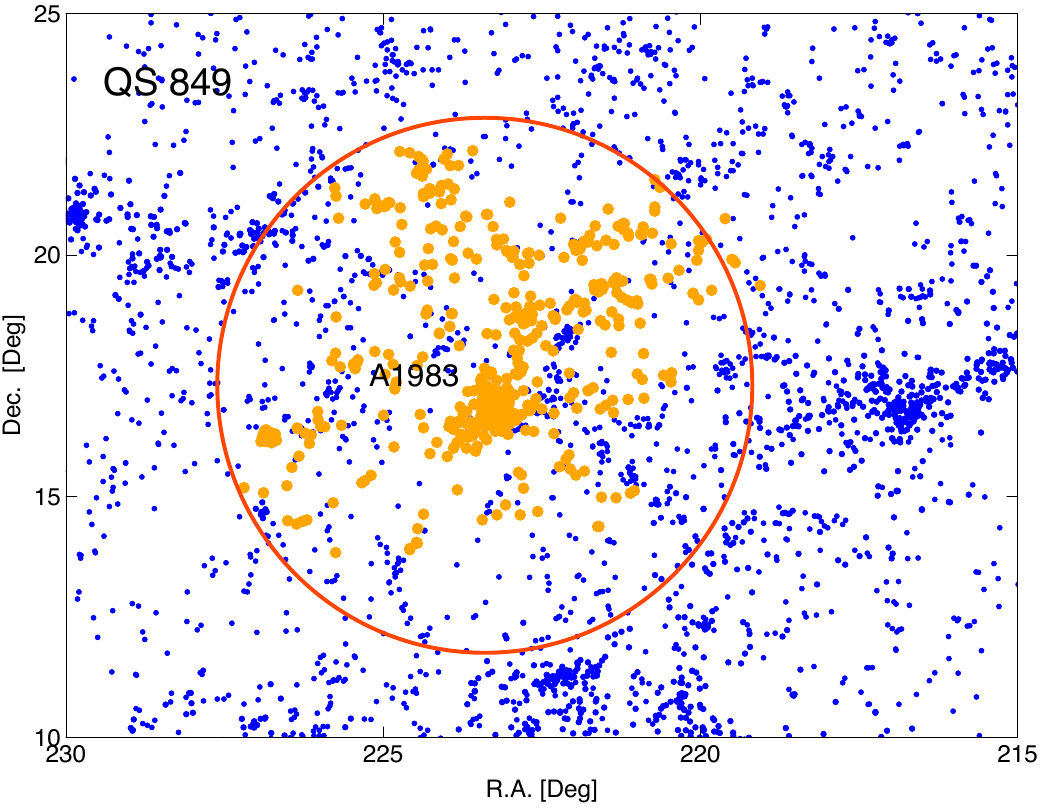}} 
\end{minipage}\\

\vspace*{0cm}

\begin{minipage}{0.45\textwidth}
\resizebox{8.0cm}{!}{\includegraphics[width = 1.6in,angle=0]{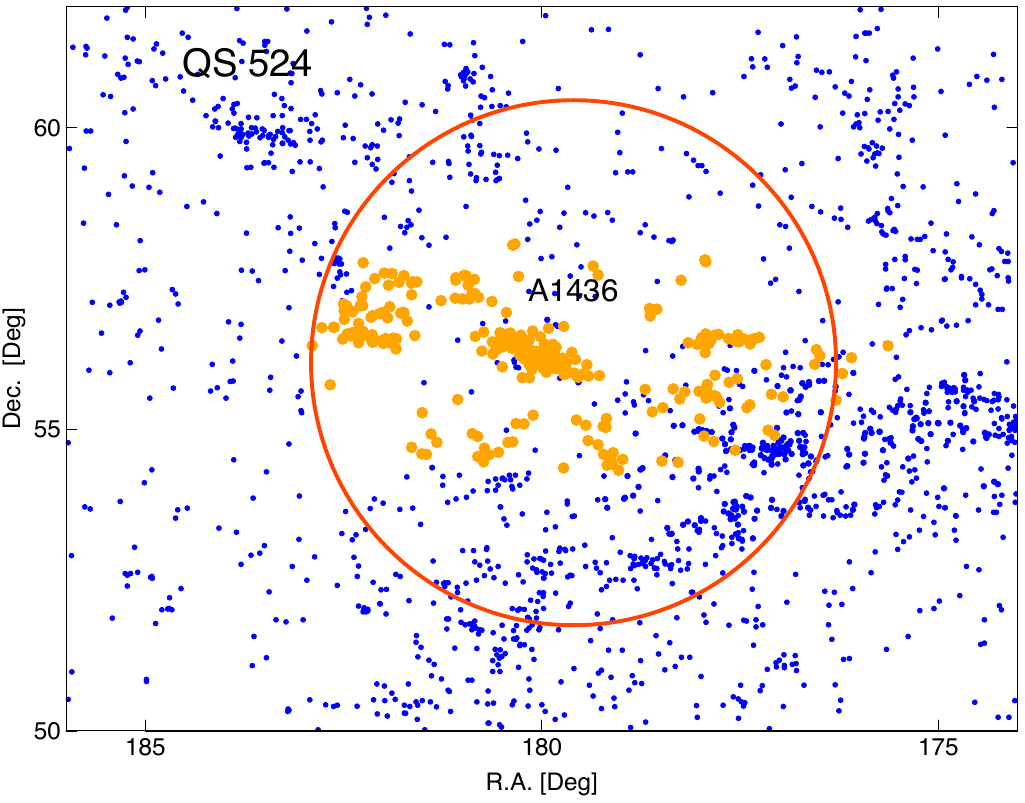}}
\end{minipage}
\begin{minipage}{0.45\textwidth}
\resizebox{8.0cm}{!}{\includegraphics[width = 1.6in,angle=0]{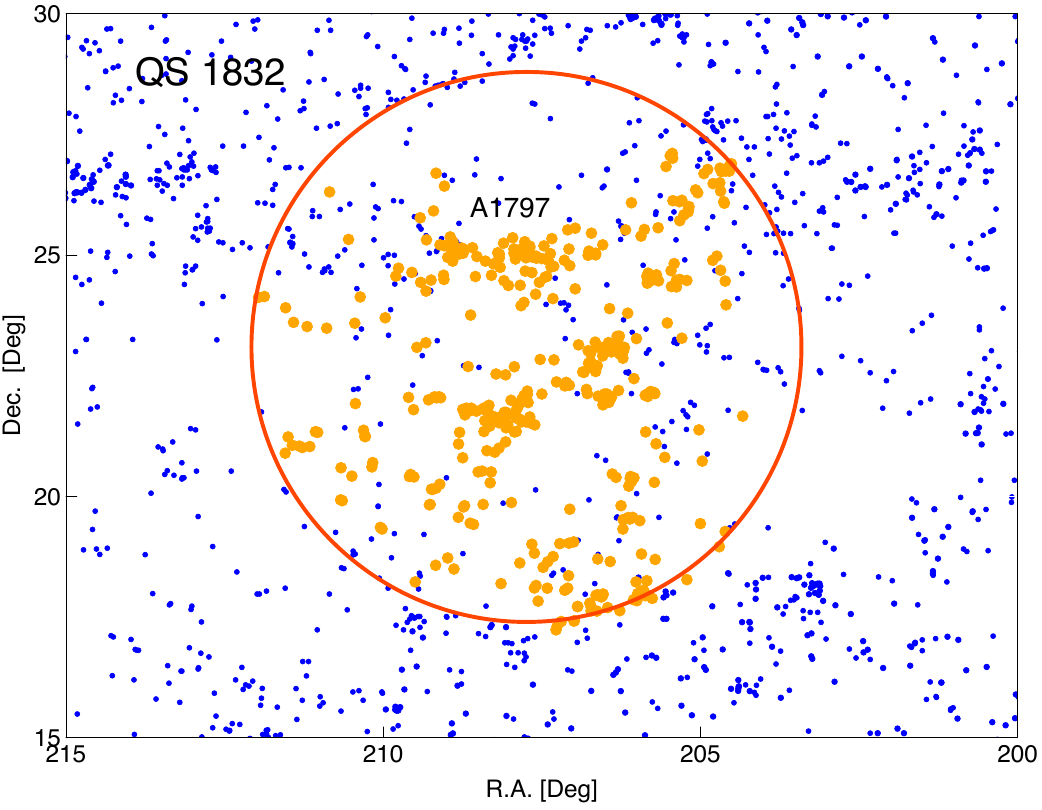}}
\end{minipage}\\
\caption {Distribution of galaxies in the regions of superclusters QS 550, QS 849, QS 524, and QS 1832 in the plane of the sky. Orange dots correspond to supercluster member galaxies and blue dots show the galaxies outside the  superclusters. A red spherical circle illustrates the spherical diameter of a supercluster (see table\ref{table1}). Abell clusters are labeled.}
\label{Figmap}
\end{figure*} 

\end{appendix}


\begin{thebibliography}{}

\bibitem[1995]{bahcall95}
Bahcall, N.A., Lubin, L.M., \& Dorman, V. 1995, ApJ, 447, 81

\bibitem[2014]{bahcall14}
Bahcall, N.A., \& Kulier, A. 2014, MNRAS, 439, 3, 2505

\bibitem[2001]{basilakos01}
Basilakos, S., Plionis, M., \& Rowan-Robinson, M. 2001, MNRAS, 323, 47

\bibitem[2003]{basilakos}
Basilakos, S. 2003, MNRAS, 344, 602 

\bibitem[2003]{blanton}
Blanton, M.R., Hogg, D.W., Bahcall, N.A., et al. 2003, ApJ, 592,2 819

\bibitem[2012]{byrd12}
Byrd, G.G., Chernin, A.D., Teerikorpi, P., \& Valtonen, M.J. 2012, 
Paths to Dark Energy: Theory and Observation (de Gruyter, Berlin)

\bibitem[2021]{Boh21}
Böhringer, H., \& Chon, G. 2021, A\&A, 656, A144

\bibitem[1996]{carlberg}
Carlberg, R.G., Yee, H.K.C., Ellingson, E., et al. 1996, ApJ, 462, 32

\bibitem [1999]{Ceno} 
Cen, R. \& Ostriker, J. P. 1999, ApJ, 514, 1

\bibitem [2006]{cen06} 
Cen, R. \& Ostriker, J. P. 2006, ApJ, 650, 560

\bibitem[2001]{chernin01}
Chernin, A.D. 2001, Physics-Uspekhi, 44, 1099

\bibitem[2006]{chernin06}
Chernin A.D., Teerikorpi P., \& Baryshev Yu.V. 2006, A\&A, 456, 13

\bibitem[2012]{chernin12}
Chernin, A.D., Teerikorpi, P., Valtonen, M.J., et al. 2012, A\&A, 539, A4

\bibitem[2014]{chon14}
Chon, G., Böhringer, H., Collins, C.A., Krause, M. 2014, A\&A, 567, 144

\bibitem[2015]{chon15}
Chon, G., B\"ohringer, H., \& Zaroubi, S. 2015, A\&A, 575, L14 

\bibitem[2011]{costa}
Costa-Duarte, M.V., Sodr\'e, L.Jr., \& Durret, F. 2011, MNRAS, 411, 1716 

\bibitem[1982]{davis82} 
Davis, M., Huchra, J., Lathman, D.W., \& Tonry, J. 1982, ApJ, 253, 423

\bibitem[2019]{dupuy} 
Dupuy, A., Courtois, H.M., \& Florent, D., et al. 2019, MNRAS, 489, L1

\bibitem[2006]{dynner06} 
D\"{u}nner, R., Araya, P.A., Meza, A., Reisenegger, A. 2006, MNRAS, 366, 803

\bibitem[1980]{einasto80} 
Einasto, J., J\~oeveer, M., \& Saar, E. 1980, MNRAS, 193, 353 

\bibitem[2001]{einasto2001}
Einasto, M., Einasto, J., Tago, E., M{\"u}ller, V., \& Andernach, H. 2001, AJ, 122, 2222

\bibitem[2003]{einasto03} 
Eiansto, J., H\"utsi, G., Einasto, M., et al. 2003, A\&A 405, 425

\bibitem[2007]{einasto07} 
Einasto, M., Saar, E., \& Liivam\"agi, L.J. 2007, A\&A, 476, 697

\bibitem[2011]{einasto2011} 
Einasto, M., Liivam\"agi, L.J., Tago, E., et al. 2011, A\&A 532, A5

\bibitem[2011]{E11} 
Einasto, M., Liivam\"agi, L. J., Saar, E., et al. 2011, A\&A  535, 36 

\bibitem[2015]{einasto15} 
Einasto, M., Gramann, M., Liivam\"agi, L.J., et al. 2015, A\&A 580, A69

\bibitem[2016] {einasto16} 
Einasto, M., Lietzen, H., Gramann, M., et al. 2016, A\&A, 595, A70

\bibitem[2020]{einasto20} 
Einasto, M., Deshev, B., Tenjes, P., et al. 2020, A\&A, 641, 172

\bibitem[2021]{einasto21} 
Einasto, M., Kipper, R., Tenjes, P., et al. 2021, A\&A, 649, 51

\bibitem[2022]{einasto22} 
Einasto, M., Tenjes, P., Gramann, M., et al. 2022, 2022arXiv220408918E 

\bibitem[2014]{erdogdu04} 
Erdogdu, P., Lahav, O., Zaroubi, S. et al. 2004, MNRAS, 352, 939 

\bibitem[2008]{frieman08} 
Frieman, J.A., Turner, M.S., \& Huterer, D. 2008, ARAA, 46, 385 

\bibitem[2004]{gavazzi}
Gavazzi, R., Mellier, Y., Fort, B., Cuillandre, J.-C., \& Dantel-Fort, M. 2004, A\&A, 422, 407


\bibitem[2015]{gramann15}
Gramann, M., Einasto, M., Hein\"am\"aki, P., et al. 2015, A\&A, 581, 135

\bibitem[2018]{haider16}
Haider, M., Steinhauser, D., Vogelsberger, M., et al. 2016, MNRAS, 457, 3024

\bibitem[2018]{hofmeister18}
Hofmeister, A.M., Criss, R.E., \& Criss, E.M. 2018, Planetary \& Space Science, 152, 68

\bibitem[1998]{jaaniste}
Jaaniste, J., Tago, E., Einasto, M., et al. 1998, A\&A, 336, 35

\bibitem[1968]{kara68}
Karachentsev, I.D. 1968, Publ. Byurakan Obs., 39, 76

\bibitem[2012]{liivamagi12}
Liivam\"{a}gi, L.J., Tempel, E., \& Saar, E. 2012, A\&A 539, A80

\bibitem[2011]{luparello11} 
Luparello, H., Lares, M., Lambas, D.G., Padilla, N. 2011, MNRAS, 415, 964

\bibitem[19]{macmillan30}
MacMillan, W.D. 1930, The Theory of the Potential, (McGraw-Hill, New York 1930)

\bibitem[2016]{nadathur16}
Nadathur, S., \& Crittenden, R. 2016, ApJL, 830. L10  

\bibitem[2014]{Old13}
Old, L., Skibba, R.A., Pearce, F.R, et al. 2014, MNRAS, 441, 2, 1513

\bibitem[2015]{Old15}
Old, L., Wojtak, R., Mamon, G.A, et al. 2015, MNRAS, 449, 1897

\bibitem[2014]{pearson14}
Pearson, D.W., Batiste, M., \& Batuski, D.J. 2014, MNRAS, 441, 1601

\bibitem[2014]{planck14}
Planck Collaboration: Ade, P.A.R., Aghanim, N., Armitage-Caplan, C., et al. 2014, A\&A, 571, A16

\bibitem[2016]{planckymap}
Planck Collaboration: Aghanim, N., Arnaud, M., Ashdown, M., et al. 2016, A\&A, 594, A22

\bibitem[2006]{Ragone}
Ragone, C.J., Muriel, H., et al. 2006, A\&A, 445, 819

\bibitem[2007]{saar2007}
 Saar, E., Martínez, V.J., Starck, J.-L., \& Donoho, D.L. 2007, MNRAS, 374, 1030 

\bibitem[2004]{shandarin}
Shandarin, S.F., Sheth, J.V., \& Sahni, V. 2004, MNRAS, 353, 162

\bibitem[1998]{sahni} 
Sahni, V., Sathyaprakash, B.S., \& Shandarin, S.F. 1998, ApJ, 495, L5

\bibitem[2011]{schirmer}
Schirmer, M., Hildebrandt, H., Kuijken, K., \& Erben, T. 2011, A\&A, 532, 57

\bibitem[2011]{Sheth}
Sheth, R., \&  Diaferio, A. 2011, A\&A, MNRAS, 417, 4

\bibitem[2010]{tago10}
Tago, E., Saar, E., Tempel, E., et al. 2010, A\&A, 514, A102

\bibitem[2010]{teerikorpi10}
Teerikorpi, P., \& Chernin A.D. 2010, A\&A, 516, 93

\bibitem[2015]{teerikorpi15}
Teerikorpi, P., Hein\"{a}m\"{a}ki, P.,  Nurmi, P., et al. 2015, A\&A, 577, A144


\bibitem[2018]{teeri18}
Teerikorpi, P.,  Hein\"{a}m\"{a}ki, P., \& Chernin, A.D. 2018, AN, 339, 705 

\bibitem[2011]{tempel11}
Tempel, E., Saar, E., Liivam\"agi, L.J., et al. 2011, A\&A, 529, A53

\bibitem[2012]{tempel12}
Tempel, E., Tago, E., Liivam\"agi, L.J., et al. 2012, A\&A, 550, A106

\bibitem[2014]{tempel14}
Tempel, E., Tamm, A., Gramann, M., et al. 2014, A\&A, 566, A1

\bibitem[2014]{tully14}
Tully, R. B., Courtois, H., Hoffman, Y., \& Pomar\`{e}de, D. 2014, Nature, 513, 71 

\bibitem[1970]{vauco70}
Vaucouleurs, G. de. 1970, Science, 167, 1203

\bibitem[1967]{woltjer67}
Woltjer, L. 1967, in Lectures in Applied Mathematics (American Mathematical
Society), 9, 1


\end{thebibliography}
\end{document}